\pdfoutput=1

\documentclass[12pt,a4paper]{article}

\usepackage{multirow}
\usepackage{rotating}

\usepackage{ifthen} 
\newboolean{pdflatex}
\setboolean{pdflatex}{true} 

\newboolean{articletitles}
\setboolean{articletitles}{true} 

\newboolean{uprightparticles}
\setboolean{uprightparticles}{false} 

\newboolean{inbibliography}
\setboolean{inbibliography}{false} 

\usepackage{microtype}
\usepackage{lineno}  
\usepackage{xspace} 
\usepackage{caption} 

\usepackage{graphicx}  
\usepackage{color}
\usepackage{colortbl}
\graphicspath{{./figs/}{./figs-S/}} 

\usepackage{amsmath} 
\usepackage{amssymb}
\usepackage{amsfonts}
\usepackage{upgreek} 

\newcommand*\patchAmsMathEnvironmentForLineno[1]{%
\expandafter\let\csname old#1\expandafter\endcsname\csname #1\endcsname
\expandafter\let\csname oldend#1\expandafter\endcsname\csname
end#1\endcsname
 \renewenvironment{#1}%
   {\linenomath\csname old#1\endcsname}%
   {\csname oldend#1\endcsname\endlinenomath}%
}
\newcommand*\patchBothAmsMathEnvironmentsForLineno[1]{%
  \patchAmsMathEnvironmentForLineno{#1}%
  \patchAmsMathEnvironmentForLineno{#1*}%
}
\AtBeginDocument{%
\patchBothAmsMathEnvironmentsForLineno{equation}%
\patchBothAmsMathEnvironmentsForLineno{align}%
\patchBothAmsMathEnvironmentsForLineno{flalign}%
\patchBothAmsMathEnvironmentsForLineno{alignat}%
\patchBothAmsMathEnvironmentsForLineno{gather}%
\patchBothAmsMathEnvironmentsForLineno{multline}%
\patchBothAmsMathEnvironmentsForLineno{eqnarray}%
}

\usepackage{hyperref}    
\usepackage[all]{hypcap} 

\def\lhcb {\mbox{LHCb}\xspace}
\def\atlas  {\mbox{ATLAS}\xspace}
\def\cms    {\mbox{CMS}\xspace}

\def\lhc    {\mbox{LHC}\xspace}

\ifthenelse{\boolean{uprightparticles}}%
{

 \def\Pmu         {\ensuremath{\upmu}\xspace}                 
 \def\Pnu         {\ensuremath{\upnu}\xspace}

 \def\PDelta      {\ensuremath{\Delta}\xspace}                 
 \def\PXi      {\ensuremath{\Xi}\xspace}                 
 \def\PLambda      {\ensuremath{\Lambda}\xspace}                 
 \def\PSigma      {\ensuremath{\Sigma}\xspace}                 
 \def\POmega      {\ensuremath{\Omega}\xspace}                 
 \def\PUpsilon      {\ensuremath{\Upsilon}\xspace}                 
 
 \def\PB      {\ensuremath{\mathrm{B}}\xspace}                 
                  
 \def\PD      {\ensuremath{\mathrm{D}}\xspace}

 \def\PK      {\ensuremath{\mathrm{K}}\xspace}

 \def\PW      {\ensuremath{\mathrm{W}}\xspace}

 \def\PZ      {\ensuremath{\mathrm{Z}}\xspace}                 
                  
 \def\Pb      {\ensuremath{\mathrm{b}}\xspace}                 
 \def\Pc      {\ensuremath{\mathrm{c}}\xspace}

 \def\Pi      {\ensuremath{\mathrm{i}}\xspace}

 \def\Pp      {\ensuremath{\mathrm{p}}\xspace}

}
{

 \def\Pmu         {\ensuremath{\mu}\xspace}                 
 \def\Pnu         {\ensuremath{\nu}\xspace}

 \mathchardef\PDelta="7101
 \mathchardef\PXi="7104
 \mathchardef\PLambda="7103
 \mathchardef\PSigma="7106
 \mathchardef\POmega="710A
 \mathchardef\PUpsilon="7107
                  
 \def\PB      {\ensuremath{B}\xspace}                 
                  
 \def\PD      {\ensuremath{D}\xspace}

 \def\PK      {\ensuremath{K}\xspace}

 \def\PW      {\ensuremath{W}\xspace}

 \def\PZ      {\ensuremath{Z}\xspace}                 
                  
 \def\Pb      {\ensuremath{b}\xspace}                 
 \def\Pc      {\ensuremath{c}\xspace}

 \def\Pi      {\ensuremath{i}\xspace}

 \def\Pp      {\ensuremath{p}\xspace}

}

\makeatletter
\ifcase \@ptsize \relax
  \newcommand{\miniscule}{\@setfontsize\miniscule{4}{5}}
\or
  \newcommand{\miniscule}{\@setfontsize\miniscule{5}{6}}
\or
  \newcommand{\miniscule}{\@setfontsize\miniscule{5}{6}}
\fi
\makeatother

\DeclareRobustCommand{\optbar}[1]{\shortstack{{\miniscule (\rule[.5ex]{1.25em}{.18mm})}
  \\ [-.7ex] $#1$}}

\def\mup        {{\ensuremath{\Pmu^+}}\xspace}
\def\mun        {{\ensuremath{\Pmu^-}}\xspace} 

\def\neu        {{\ensuremath{\Pnu}}\xspace}
\def\neub       {{\ensuremath{\overline{\Pnu}}}\xspace}

\def\W      {{\ensuremath{\PW}}\xspace}
\def\Wp     {{\ensuremath{\PW^+}}\xspace}
\def\Wm     {{\ensuremath{\PW^-}}\xspace}
\def\Wpm    {{\ensuremath{\PW^\pm}}\xspace}
\def\Z      {{\ensuremath{\PZ}}\xspace}

\def\cquark    {{\ensuremath{\Pc}}\xspace}
\def\cquarkbar {{\ensuremath{\overline \cquark}}\xspace}
\def\ccbar     {{\ensuremath{\cquark\cquarkbar}}\xspace}
\def\bquark    {{\ensuremath{\Pb}}\xspace}
\def\bquarkbar {{\ensuremath{\overline \bquark}}\xspace}
\def\bbbar     {{\ensuremath{\bquark\bquarkbar}}\xspace}

\def\kaon    {{\ensuremath{\PK}}\xspace}
  \def\Kbar    {{\kern 0.2em\overline{\kern -0.2em \PK}{}}\xspace}

\def\KorKbar    {\kern 0.18em\optbar{\kern -0.18em K}{}\xspace}

  \def\Dbar    {{\kern 0.2em\overline{\kern -0.2em \PD}{}}\xspace}

\def\DorDbar    {\kern 0.18em\optbar{\kern -0.18em D}{}\xspace}

\def\Bbar    {{\ensuremath{\kern 0.18em\overline{\kern -0.18em \PB}{}}}\xspace}

\def\BorBbar    {\kern 0.18em\optbar{\kern -0.18em B}{}\xspace}

  \def\Y#1S{\ensuremath{\PUpsilon{(#1S)}}\xspace}

\def\proton      {{\ensuremath{\Pp}}\xspace}

\def\Lbar        {{\ensuremath{\kern 0.1em\overline{\kern -0.1em\PLambda}}}\xspace}
\def\LorLbar    {\kern 0.18em\optbar{\kern -0.18em \PLambda}{}\xspace}

\def\to                 {\ensuremath{\rightarrow}\xspace}

\def\ordalsq {{\ensuremath{\mathcal{O}(\alpha^{2})}}\xspace}

\def\AT#1     {\ensuremath{A_{\mathrm{T}}^{#1}}\xspace}           

\def\C#1      {\ensuremath{\mathcal{C}_{#1}}\xspace}                       
\def\Cp#1     {\ensuremath{\mathcal{C}_{#1}^{'}}\xspace}                    
\def\Ceff#1   {\ensuremath{\mathcal{C}_{#1}^{\mathrm{(eff)}}}\xspace}        
\def\Cpeff#1  {\ensuremath{\mathcal{C}_{#1}^{'\mathrm{(eff)}}}\xspace}       
\def\Ope#1    {\ensuremath{\mathcal{O}_{#1}}\xspace}                       
\def\Opep#1   {\ensuremath{\mathcal{O}_{#1}^{'}}\xspace}                    

\newcommand{\tev}{\ifthenelse{\boolean{inbibliography}}{\ensuremath{~T\kern -0.05em eV}\xspace}{\ensuremath{\mathrm{\,Te\kern -0.1em V}}}\xspace}
\newcommand{\gev}{\ensuremath{\mathrm{\,Ge\kern -0.1em V}}\xspace}
\newcommand{\mev}{\ensuremath{\mathrm{\,Me\kern -0.1em V}}\xspace}
\newcommand{\kev}{\ensuremath{\mathrm{\,ke\kern -0.1em V}}\xspace}
\newcommand{\ev}{\ensuremath{\mathrm{\,e\kern -0.1em V}}\xspace}
\newcommand{\gevc}{\ensuremath{{\mathrm{\,Ge\kern -0.1em V\!/}c}}\xspace}
\newcommand{\mevc}{\ensuremath{{\mathrm{\,Me\kern -0.1em V\!/}c}}\xspace}
\newcommand{\gevcc}{\ensuremath{{\mathrm{\,Ge\kern -0.1em V\!/}c^2}}\xspace}
\newcommand{\gevgevcccc}{\ensuremath{{\mathrm{\,Ge\kern -0.1em V^2\!/}c^4}}\xspace}
\newcommand{\mevcc}{\ensuremath{{\mathrm{\,Me\kern -0.1em V\!/}c^2}}\xspace}

\def\mum  {\ensuremath{{\,\upmu\rm m}}\xspace}

\def\pb {\ensuremath{\rm \,pb}\xspace}
\def\invpb {\ensuremath{\mbox{\,pb}^{-1}}\xspace}

\def\invfb   {\ensuremath{\mbox{\,fb}^{-1}}\xspace}

\newcommand{\chisqndf}{\ensuremath{\chi^2/\mathrm{ndf}}\xspace}

\def\gsim{{~\raise.15em\hbox{$>$}\kern-.85em
          \lower.35em\hbox{$\sim$}~}\xspace}
\def\lsim{{~\raise.15em\hbox{$<$}\kern-.85em
          \lower.35em\hbox{$\sim$}~}\xspace}

\def\sqs   {\ensuremath{\protect\sqrt{s}}\xspace}

\def\ptot       {\mbox{$p$}\xspace}
\def\pt         {\mbox{$p_{\rm T}$}\xspace}

\def\rad{\ensuremath{\rm \,rad}\xspace}

\def\evtgen     {\mbox{\textsc{EvtGen}}\xspace}
\def\fewz       {\mbox{\textsc{Fewz}}\xspace}

\def\geant      {\mbox{\textsc{Geant4}}\xspace}

\def\photos     {\mbox{\textsc{Photos}}\xspace}

\def\pythia     {\mbox{\textsc{Pythia}}\xspace}
\def\resbos     {\mbox{\textsc{ResBos}}\xspace}

\def\tell1  {TELL1\xspace}
\def\ukl1   {UKL1\xspace}

\usepackage{cite} 
\usepackage{mciteplus}

\newcommand{\dataset}{\ensuremath{975 \pm 17\invpb}\xspace}
\newcommand{\csp}{\ensuremath{861.0 \pm 2.0 \pm 11.2 \pm 14.7 \pb}\xspace}
\newcommand{\csm}{\ensuremath{675.8 \pm 1.9 \pm 8.8 \pm 11.6 \pb}\xspace}
\newcommand{\csr}{\ensuremath{1.274 \pm 0.005 \pm 0.009}\xspace}

\newcommand{\wmn}{\ensuremath{\W \to \mu\neu}\xspace}
\newcommand{\wpmn}{\ensuremath{\Wp \to \mup\neu}\xspace}
\newcommand{\wmmn}{\ensuremath{\Wm \to \mun\neub}\xspace}
\newcommand{\wpmmn}{\ensuremath{\Wpm \to \mupm\neu}\xspace}
\newcommand{\wtn}{\ensuremath{\W \to \tau\neu}\xspace}
\newcommand{\zmm}{\ensuremath{\Z \to \mu\mu}\xspace}
\newcommand{\ztt}{\ensuremath{\Z \to \tau\tau}\xspace}

\newcommand{\pw}{{\textrm{pseudo}-\W}\xspace}

\def\mupm{\ensuremath{\Pmu^\pm}\xspace} 
\def\pp{\ensuremath{\proton\proton}\xspace}

\newcommand{\cswp}{\ensuremath{\sigma_{\wpmn}}\xspace}
\newcommand{\cswm}{\ensuremath{\sigma_{\wmmn}}\xspace}
\newcommand{\cswpm}{\ensuremath{\sigma_{\wpmmn}}\xspace}

\newcommand{\ratio}{\ensuremath{R_{\W}}\xspace}
\newcommand{\asy}{\ensuremath{A_{\mu}}\xspace}
\newcommand{\pdf}{\textsc{PDF}\xspace}
\newcommand{\pdfs}{\textsc{PDF}s\xspace}
\newcommand{\nnlo}{\textsc{NNLO}\xspace}

\newcommand{\qcd}{\textsc{QCD}\xspace}
\newcommand{\qed}{\textsc{QED}\xspace}
\newcommand{\fsr}{\textsc{FSR}\xspace}
\newcommand{\abm}{\textsc{ABM12}\xspace}

\newcommand{\ct}{\textsc{CT10}\xspace}
\newcommand{\hera}{\textsc{HERA1.5}\xspace}
\newcommand{\jr}{\textsc{JR09}\xspace}
\newcommand{\mstw}{\textsc{MSTW08}\xspace}
\newcommand{\nnpdf}{\textsc{NNPDF2.3}\xspace}

\newcommand{\ip}{\ensuremath{\textrm{IP}}\xspace}
\newcommand{\ecalo}{\ensuremath{E_{\textrm{calo}}}\xspace}
\newcommand{\eop}{\ensuremath{\ecalo / \ptot c}\xspace}
\newcommand{\rcone}{\ensuremath{R = \sqrt{\Delta\eta^{2} + \Delta\phi^{2}}}\xspace}
\newcommand{\ptcone}{\ensuremath{p_{\textrm{T}}^{\textrm{cone}}}\xspace}
\newcommand{\etcone}{\ensuremath{E_{\textrm{T}}^{\textrm{cone}}}\xspace}
\newcommand{\mmm}{\ensuremath{m_{\mu\mu}}\xspace}
\newcommand{\nc}{\ensuremath{N_{W}}\xspace}
\newcommand{\effrec}{\ensuremath{\varepsilon_{\textrm{rec}}}\xspace}
\newcommand{\effsel}{\ensuremath{\varepsilon_{\textrm{sel}}}\xspace}
\newcommand{\acc}{\ensuremath{\mathcal{A}^{\pm}}\xspace}
\newcommand{\ffsr}{\ensuremath{f^{\pm}_{\textrm{\fsr}}}\xspace}

\newcommand{\lumi}{\ensuremath{\mathcal{L}}\xspace}

\begin{document}

\renewcommand{\thefootnote}{\fnsymbol{footnote}}
\setcounter{footnote}{1}


\begin{titlepage}
\pagenumbering{roman}

\vspace*{-1.5cm}
\centerline{\large EUROPEAN ORGANIZATION FOR NUCLEAR RESEARCH (CERN)}
\vspace*{1.5cm}
\hspace*{-0.5cm}
\begin{tabular*}{\linewidth}{lc@{\extracolsep{\fill}}r}
\ifthenelse{\boolean{pdflatex}}
{\vspace*{-2.7cm}\mbox{\!\!\!\includegraphics[width=.14\textwidth]{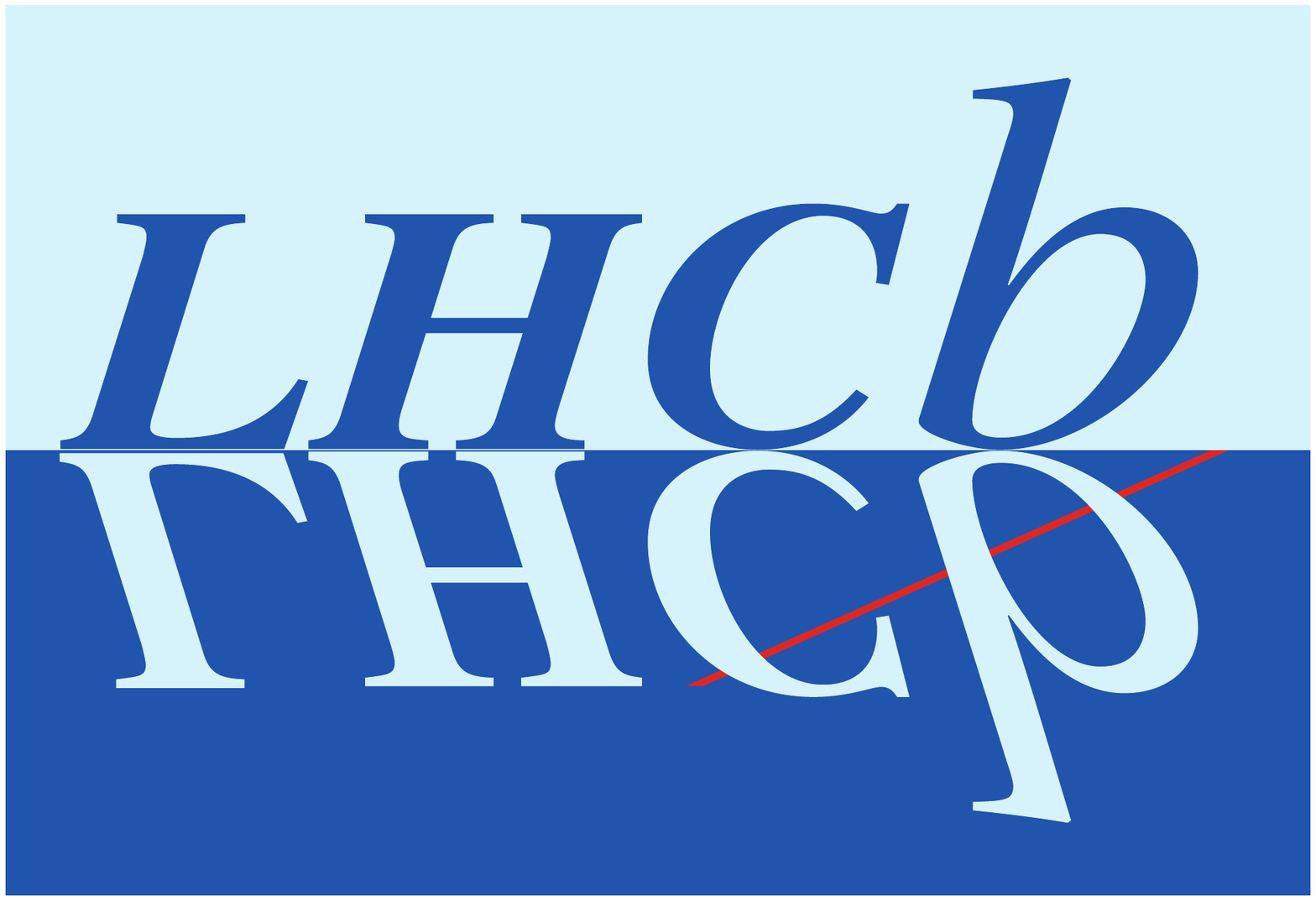}} & &}%
{\vspace*{-1.2cm}\mbox{\!\!\!\includegraphics[width=.12\textwidth]{lhcb-logo.eps}} & &}%
\\
 & & CERN-PH-EP-2014-175 \\
 & & LHCb-PAPER-2014-033 \\
 & & December 18, 2014 \\
\end{tabular*}

\vspace*{0.75cm}

{\bf\boldmath\huge\begin{center}
Measurement of the forward \\ \W boson cross-section \\ in \pp collisions at \sqs = 7\tev
\end{center}}

\vspace*{0.5cm}

\begin{center}
The LHCb collaboration\footnote{Authors are listed at the end of this paper.}
\end{center}

\vspace*{0.5cm}

\begin{abstract}
\noindent A measurement of the inclusive \wmn production cross-section using data from \pp collisions at a centre-of-mass energy of \sqs = 7\tev is presented. The analysis is based on an integrated luminosity of about $1.0\invfb$ recorded with the \lhcb detector. Results are reported for muons with a transverse momentum greater than 20\gevc and pseudorapidity between 2.0 and 4.5. The \Wp and \Wm production cross-sections are measured to be
\begin{align*}
\cswp &= \csp, \\
\cswm &= \csm,
\end{align*}
\noindent where the first uncertainty is statistical, the second is systematic and the third is due to the luminosity determination. Cross-section ratios and differential distributions as functions of the muon pseudorapidity are also presented. The ratio of \Wp to \Wm cross-sections in the same fiducial kinematic region is determined to be
\begin{equation*}
\frac{\cswp}{\cswm} = \csr,
\end{equation*}
\noindent where the uncertainties are statistical and systematic, respectively. Results are in good agreement with theoretical predictions at next-to-next-to-leading order in perturbative quantum chromodynamics.
\end{abstract}

\vspace*{0.5cm}

\begin{center}
Published in JHEP 12 (2014) 079
\end{center}

\vspace*{0.25cm}

{\footnotesize 
\centerline{\copyright~CERN on behalf of the \lhcb collaboration, license \href{http://creativecommons.org/licenses/by/4.0/}{CC-BY-4.0}.}}
\vspace*{0.25cm}

\end{titlepage}

\newpage
\setcounter{page}{2}
\mbox{~}
\newpage

\cleardoublepage

\renewcommand{\thefootnote}{\arabic{footnote}}
\setcounter{footnote}{0}

\pagestyle{plain} 
\setcounter{page}{1}
\pagenumbering{arabic}


\section{Introduction} \label{sec:Introduction}

Measurements of the production cross-section of electroweak bosons constitute an important test of the Standard Model at \lhc energies. Theoretical predictions, which are available at next-to-next-to-leading order (\nnlo) in perturbative quantum chromodynamics (\qcd)~\cite{DY-NNLO1, DY-NNLO2, DY-NNLO3, DY-NNLO4, DY-NNLO5}, rely on the parameterisations of the momentum fraction, Bjorken-$x$, of the partons inside the proton. The uncertainties on the parton density functions (\pdfs) of the proton are larger at low Bjorken-$x$ values. The forward acceptance of the \lhcb experiment, with a pseudorapidity coverage in the range $2 < \eta < 5$, can directly access this region of the phase space to Bjorken-$x$ values as low as $10^{-6}$ and provide constraints on the \pdfs that complement those of the  \atlas and \cms experiments~\cite{PDF-MSTW}. In addition to the determination of the \W boson cross-section, the ratio of \Wp to \Wm cross-sections and the \W production charge asymmetry allow the Standard Model to be tested with high precision, as experimental and theoretical uncertainties partially cancel.

\lhcb has previously reported measurements of inclusive \W and \Z boson production in \pp collisions at a centre-of-mass energy of \sqs = 7\tev to final states containing muons using a data sample corresponding to an integrated luminosity of about 40\invpb~\cite{LHCb-PAPER-2012-008}. The impact of the \W results on the \pdf determination was studied and determined to be the largest compared to other \lhc Drell-Yan data due to the forward acceptance of the \lhcb detector~\cite{PDF-ABM12, PDF-NNPDF23}. A more precise measurement of the \W production can therefore further improve the knowledge of the partonic content of the proton. The analysis presented in this paper is based on a data set collected in 2011, which corresponds to an integrated luminosity of approximately 1.0\invfb. The contribution of virtual photons to \Z boson production is always implied. The inclusive \Wp and \Wm cross-sections, \cswpm, are measured for muons with a transverse momentum, \pt, greater than 20\gevc and $\eta$ between 2.0 and 4.5. The differential cross-sections, the \Wp to \Wm cross-section ratio, \ratio, and the muon charge asymmetry, \asy, are determined in the same fiducial kinematic region in eight bins of muon pseudorapidity. The \atlas~\cite{ATLAS} and \cms~\cite{CMS} collaborations already reported measurements of these quantities, although in different kinematic volumes. Measurements are corrected for final state radiation (\fsr) of the muon. No corrections are applied for initial state radiation, electroweak effects, and their interplay with \qed effects.

The paper is organised as follows: Sect.~\ref{sec:Detector} describes the \lhcb detector as well as the data and the simulation samples used; the selection of the \wmn candidates and the extraction of the signal yield are outlined in Sects.~\ref{sec:Selection} and \ref{sec:Yield}; Sect.~\ref{sec:Measurement} describes the cross-section measurement and discusses the associated systematic uncertainties; the results are presented in Sect.~\ref{sec:Results}; and Sect.~\ref{sec:Conclusions} concludes the paper.

\section{Detector and data sets}\label{sec:Detector}

The \lhcb detector~\cite{Alves:2008zz} is a single-arm forward spectrometer covering the pseudorapidity range $2<\eta <5$, designed predominantly for the study of particles containing \bquark or \cquark quarks. The detector includes a high-precision tracking system consisting of a silicon-strip vertex detector surrounding the \pp interaction region, a large-area silicon-strip detector located upstream of a dipole magnet with a bending power of about $4{\rm\,Tm}$, and three stations of silicon-strip detectors and straw drift tubes placed downstream of the magnet. The magnet polarity can be reversed, so that left-right detector asymmetry effects can be studied and corrected for in the analyses. The tracking system provides a measurement of momentum, \ptot, with a relative uncertainty that varies from 0.4\% at low \ptot values to 0.6\% at 100\gevc. The minimum distance of a track to a primary vertex, the impact parameter (\ip), is measured with a resolution of $(15 + 29 / \pt) \mum$, where the \pt is in \gevc. Different types of charged hadrons are distinguished using information from two ring-imaging Cherenkov detectors. Photon, electron and hadron candidates are identified by a calorimeter system consisting of scintillating-pad and preshower detectors, an electromagnetic and a hadronic calorimeter. Muons are identified by a system composed of alternating layers of iron and multiwire proportional chambers.
The trigger consists of a hardware stage, based on information from the calorimeter and muon systems, followed by a software stage, which applies full event reconstruction. A set of global event cuts that prevents events with high occupancy dominating the processing time of the software trigger is also applied.
 
This measurement is based on data corresponding to an integrated luminosity of \dataset taken in \pp collisions at a centre-of-mass energy of 7\tev. The absolute luminosity scale was measured periodically throughout the data taking period using both Van der Meer scans~\cite{LUMI-MEER}, where the beam profile is determined by moving the beams transversely across each other, and a beam-gas imaging method~\cite{FerroLuzzi:2005em}, in which beam-gas interaction vertices near the beam crossing point are reconstructed to determine the beam profile. The two methods give consistent results and the integrated luminosity is determined from their average, with an estimated systematic uncertainty of 1.7\%~\cite{LHCB-PAPER-2014-047}.

Simulated data are used to optimise the event selection, estimate the background contamination and check the efficiencies. In the simulation, \pp collisions are generated using \pythia~\cite{Sjostrand:2006za} with a specific \lhcb configuration~\cite{LHCb-PROC-2010-056}. Decays of hadronic particles are described by \evtgen~\cite{Lange:2001uf}, in which final state radiation (\fsr) is generated using \photos~\cite{Golonka:2005pn}. The interaction of the generated particles with the detector and its response are implemented using the \geant toolkit~\cite{Allison:2006ve, Agostinelli:2002hh} as described in Ref.~\cite{LHCb-PROC-2011-006}. Additional simulated samples of \wmn and \zmm events are generated with the program \resbos~\cite{GEN-RESBOS1, GEN-RESBOS2, GEN-RESBOS3} using the \ct\nnlo~\cite{PDF-CT10} \pdf set. \resbos includes an approximate \nnlo\xspace\ordalsq calculation, plus a next-to-next-to-leading logarithm approximation for the resummation of the soft radiation at low transverse momentum of the vector boson.

Results are compared to theoretical predictions calculated with the \fewz~\cite{GEN-FEWZ2, GEN-FEWZ3} generator at \nnlo for the \abm~\cite{PDF-ABM12}, \ct~\cite{PDF-CT10}, \hera~\cite{PDF-HERA15}, \jr~\cite{PDF-JR09}, \mstw~\cite{PDF-MSTW08} and \nnpdf~\cite{PDF-NNPDF23} \nnlo \pdf sets. The scale uncertainties are estimated by independently varying the renormalisation and factorisation scales by a factor of two around the nominal value, which is set to the boson mass. The total uncertainty for each set corresponds to the \pdf (68\% CL) and the scale uncertainties added in quadrature.

\section{Event selection}\label{sec:Selection}

The signature of a \wmn decay consists of an isolated high transverse momentum muon produced at the \pp interaction point. The events must first pass a trigger decision that requires the presence of at least one muon with $\pt > 10\gevc$. Events are subsequently selected to contain at least one well reconstructed muon with a transverse momentum greater than 20\gevc and a pseudorapidity in the range $2.0 < \eta < 4.5$. These conditions define the fiducial kinematic region of the measurement. Additional requirements are imposed to discriminate signal candidates from background events.

While leptons from electroweak boson decays are expected to be isolated, hadronic \qcd processes tend to be associated with jets and produce particles with more activity around them. The muon isolation is described by using the scalar sum of the transverse momentum of all charged particles in the event excluding the candidate, \ptcone, in a cone of radius $R = 0.5$, as well as an analogous variable based on calorimeter information not associated to any track, \etcone. The cone radius is defined as \rcone, where $\Delta\eta$ ($\Delta\phi$) is the difference in pseudorapidity (azimuthal angle) between the muon and the extra particle. The \ptcone and the \etcone variables are required to be smaller than 2\gevc and  2\gev, respectively.

Drell-Yan events with two muons within the detector acceptance are removed by vetoing on the presence of a second muon with $\pt > 2\gevc$.

Owing to the long lifetime of the tau lepton, as well as of beauty and charm hadrons, muons produced in \ztt, \wtn or semileptonic decays of heavy flavour hadrons mostly originate far from the interaction point where the two protons have collided. Thus selected candidates are required to have an $\ip < 40 \mum$.

Muons typically deposit little energy in the calorimeters in contrast to high-momentum kaons or pions punching through the detector to the muon chambers. Mis-identification due to punchthrough is reduced by requiring $\eop < 4\%$, where \ecalo is the sum of the energy deposition in the electromagnetic and hadronic calorimeters associated to the particle.

In order to optimise the \wmn selection and evaluate the corresponding efficiency from data, a control sample, \pw, is selected by requiring a pair of well reconstructed tracks identified as muons with an invariant mass, \mmm, within 10\gevcc of the nominal \Z boson mass. Each of the leptons is alternately masked to mimic the presence of a neutrino and so fake two \wmn decays. A total of 47\hspace{0.5mm}503 \zmm candidates are identified. Given the stringent requirement on the reconstructed dimuon mass, the sample purity is expected to be larger than 99.7\%, as observed in Ref.~\cite{LHCb-PAPER-2012-008} within a wider mass window, and so no background subtraction is performed. The \pw sample is generally in good agreement with \wmn simulation. A small difference is observed in momentum-dependent quantities due to a larger average \pt in \zmm than \wmn decays, which is corrected using simulation.

\section{Signal yield}\label{sec:Yield}

A total of 806\hspace{0.5mm}094 \wmn candidate events fulfil the selection requirements. The signal yield is determined by simultaneously fitting the \pt spectra of positively and negatively charged muons in data to the expected shapes for signal and background contributions in eight bins of muon pseudorapidity. A template fit is performed for transverse momenta between 20 and 70\gevc using an extended maximum likelihood method based on Ref.~\cite{TFRACTIONFITTER}. The following contributions are considered, where all simulation-based templates are corrected to account for differences in the reconstruction and the selection efficiencies as observed between data and simulation.

The \wmn shape, which is defined by the \pt of the muon, is modelled using simulation by correcting reconstructed \pythia events with \resbos. The signal normalisation is allowed to vary independently in each bin of muon pseudorapidity and for each charge.

The templates of \zmm, \wtn and \ztt decays with only one muon in the \lhcb acceptance are described using simulation. In addition, \zmm events are weighted to reproduce the \resbos muon \pt distribution. The normalisation of the electroweak backgrounds is fixed by scaling the number of \Z decays with both muons in the detector acceptance, as reconstructed in the \pw data sample, by the probability of having only one muon in \lhcb instead of two, as determined from simulation. The number is corrected for differences in reconstructing and selecting \Z and \W events and takes into account acceptance variations between \wtn, \ztt and \zmm decays. The electroweak backgrounds account for $(9.82 \pm 0.16)\%$ of the total number of candidates, where the uncertainty is statistical and dominated by the size of the \pw sample.

\begin{figure}[!t]
\begin{center}
\includegraphics[width=.49\textwidth]{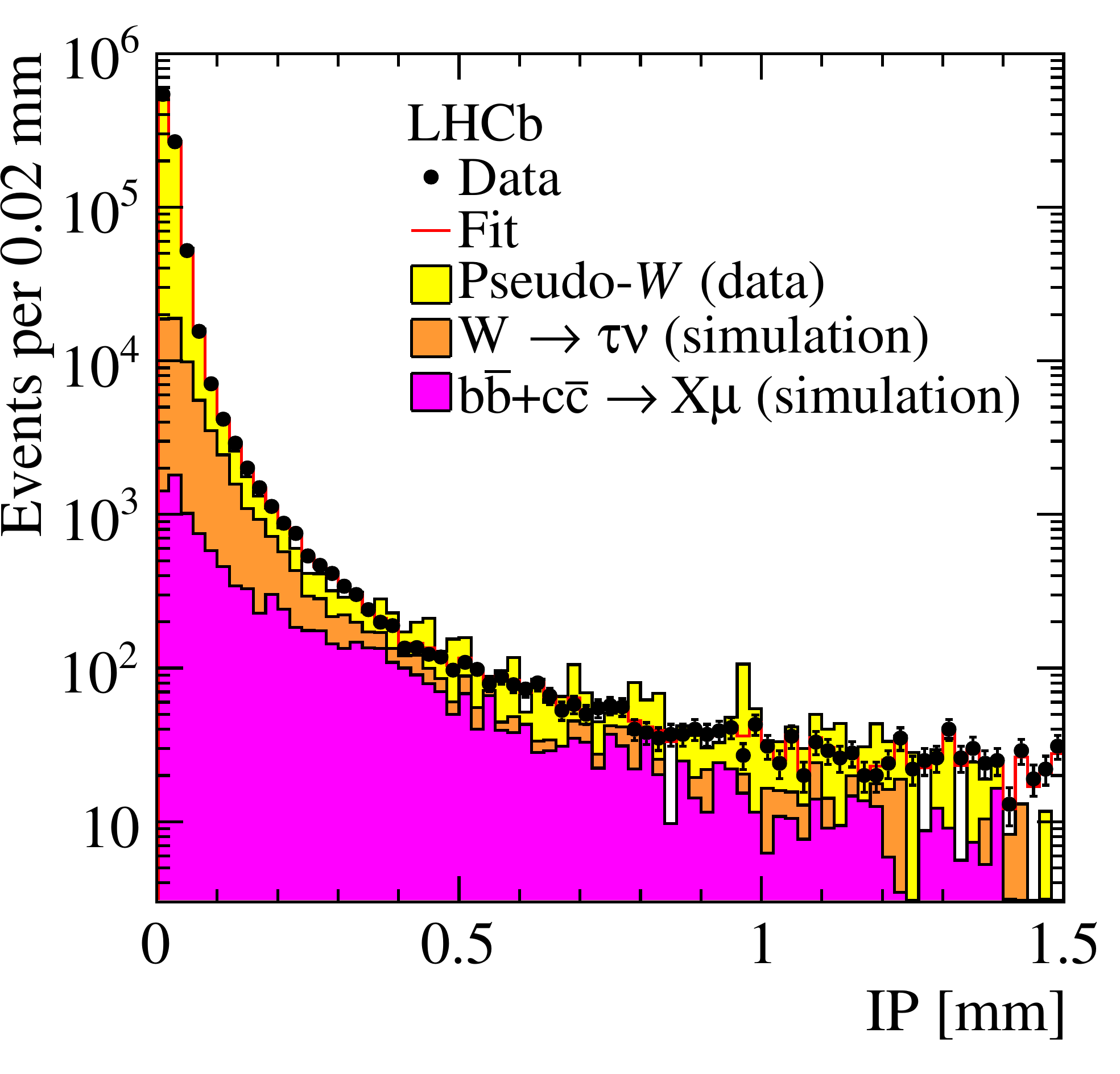}
\includegraphics[width=.49\textwidth]{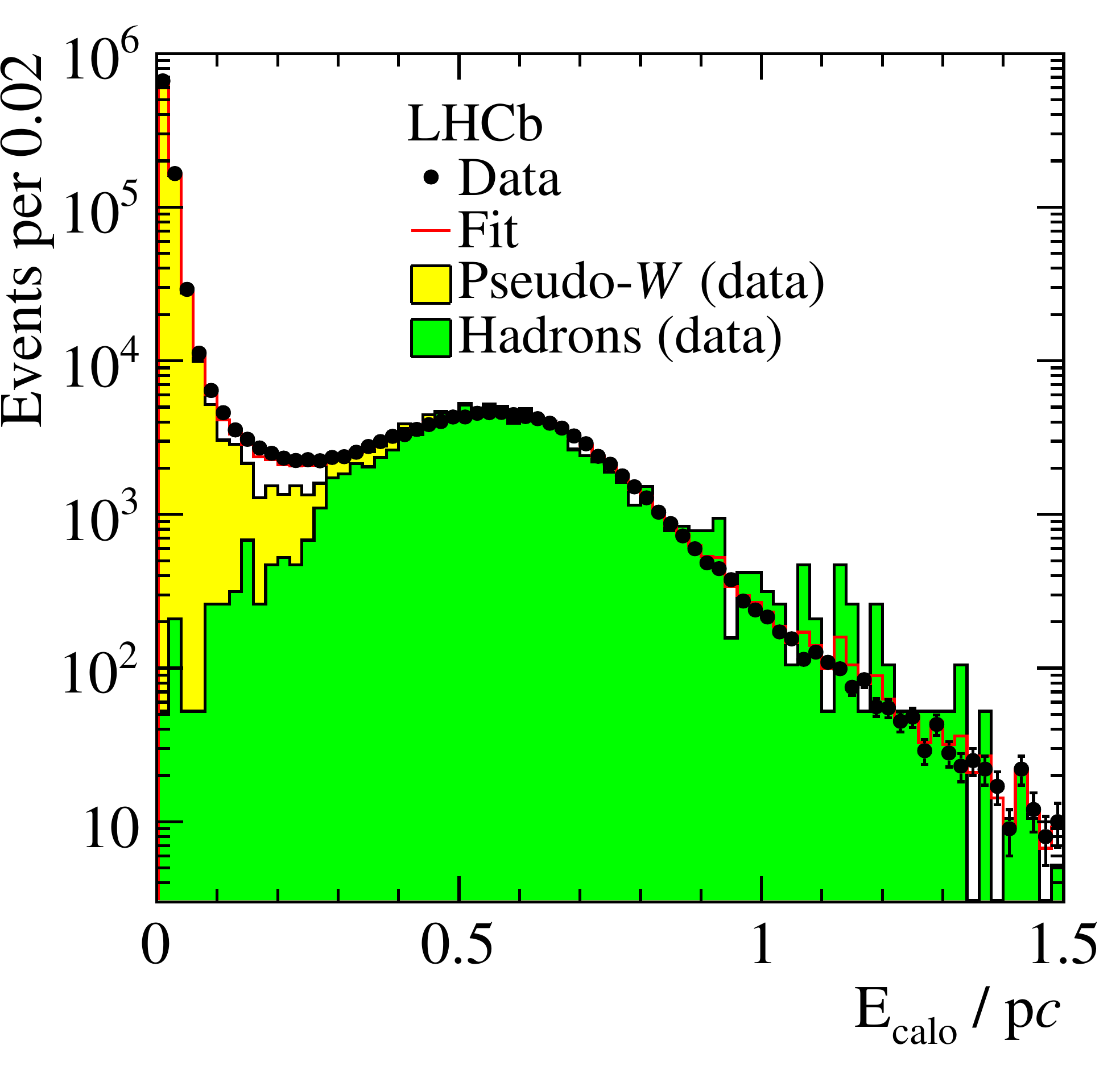}
\caption{Fit to (left) the impact parameter distribution and (right) the relative energy deposition in the calorimeters. The data are compared to fitted contributions described in the legends.}
\label{fig:bkg}
\end{center}
\end{figure}

The \pt spectrum of semileptonic decays of heavy flavour hadrons is obtained from data by requiring an impact parameter exceeding 100\mum, which selects tracks incompatible with production at a primary vertex. Several \ip thresholds are tested with marginal changes to the \pt distribution, as cross-checked on data and simulation. The normalisation is determined via a template fit to the \ip distribution in data with three classes of events: muons originating from the \pp interaction point, described with \pw data, muons coming from \wtn or \ztt decays, parameterised with simulation, and muons from heavy flavour hadrons, modelled with a mixture of inclusive \bbbar and \ccbar simulated samples. The result of the fit is shown in Fig.~\ref{fig:bkg} (left) from where the heavy flavour normalisation is determined to be $(0.48 \pm 0.03)\%$, and constrained in the \pt fit.

The residual background due to \kaon and $\pi$ punchthrough is determined from data using a template fit. Figure~\ref{fig:bkg} (right) shows the fit to the relative energy deposition in the calorimeters, where \wmn candidates are compared to muons and hadrons selected from \pw and randomly triggered data, respectively. By requiring $\eop < 4\%$, kaons or pions punching through the detector to the muon chambers are reduced to a negligible level and therefore not considered when measuring the \wmn yield. 

Kaons and pions decaying in flight to muons are described using data. Tracks that are not involved in any trigger decision are selected and the corresponding \pt is weighted by their probability to decay to muons, as measured on data. The probability is determined as the fraction of reconstructed particles identified as muons, and is found to be consistent with an estimate based on the mean lifetimes and decay distance of simulated \kaon and $\pi$ mesons. The normalisation of the decay-in-flight background is allowed to vary independently in each $\eta$ bin and for each charge.

The signal purity, defined as the ratio of signal to candidate event yield, is determined with a fit that relies on the correct description of the \pt distribution in data and thus a good calibration of the momentum scale is needed. The momentum measurement, which is based on the curvature of charged particle trajectories, critically depends on the correct alignment of the \lhcb detector and on an accurate knowledge of the magnetic field across the tracking volume. To calibrate the momentum scale for high-\pt particles, di muon pairs are identified with the same criteria used to select the \pw sample, but with \mmm in the range from 60 to 120\gevcc. The calibration is performed in bins of $\eta$ and $\phi$ for positively and negatively charged muons separately, by exploiting the deviations of the average reconstructed invariant mass from the known \Z boson mass. The procedure removes the dependence of \mmm on the direction of flight of the muon, as well as of the \pt on the charge of the particle.

The result of the fit in the range $2.0 < \eta< 4.5$ is presented in Fig.~\ref{fig:tff0}, while Figs.~\ref{fig:tff1} and \ref{fig:tff2} in Appendix~\ref{app:fit} show the fit in the eight pseudorapidity bins. The \chisqndf of the fit is 1.65, with 768 degrees of freedom. The distribution of the normalised fit residuals show an imperfect description of the data by the adopted templates, particularly for negative muons at high \pt values, however, the effect on the signal yield is small. The \wpmn and \wmmn sample purities are determined to be $(77.13 \pm 0.19)\%$ and $(77.39 \pm 0.23)\%$, respectively.

\begin{figure}[!t]
\begin{center}
\includegraphics[width=.7\textwidth]{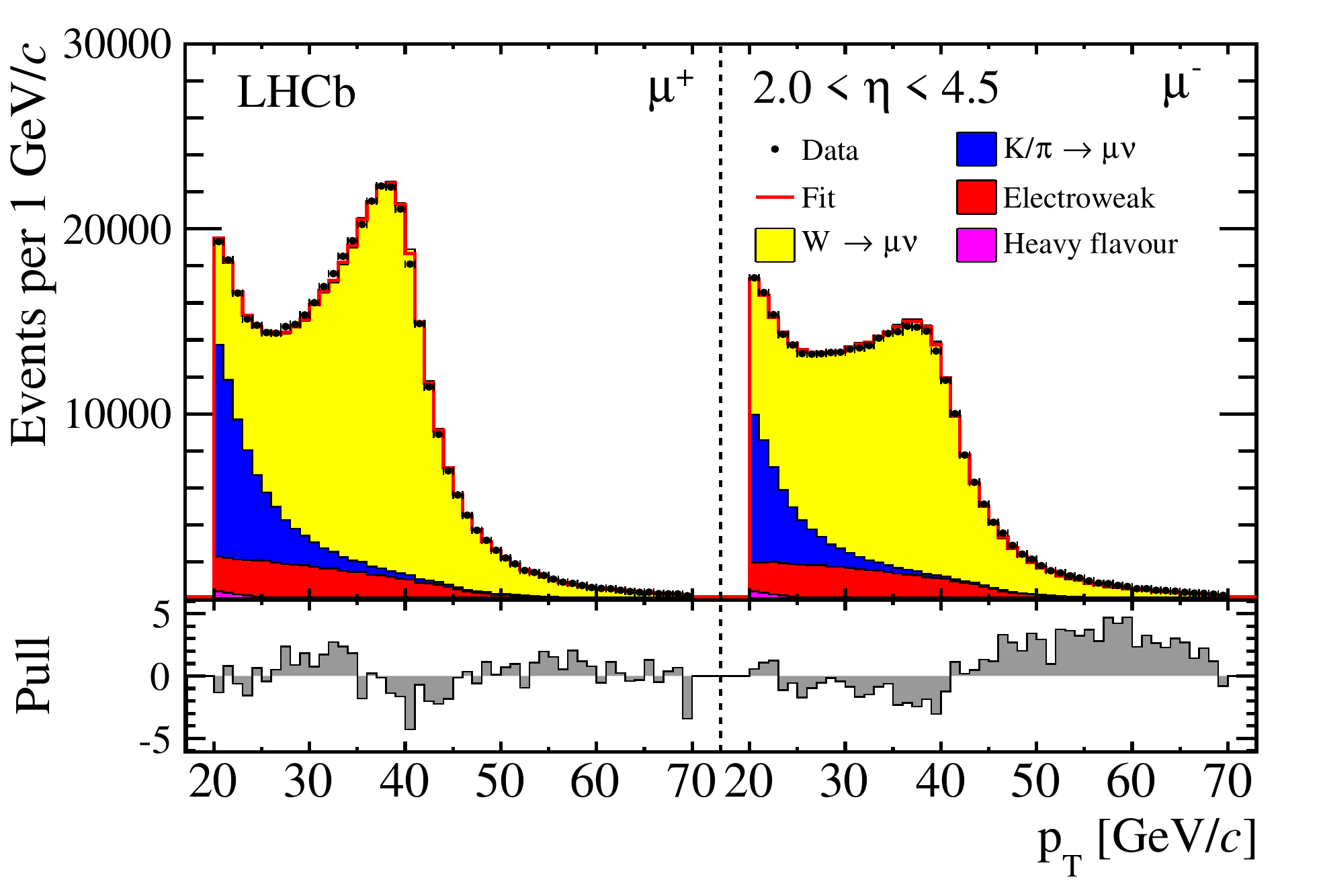}
\caption{Transverse momentum distribution of the (left panel) positive and (right panel) negative muon candidates in the fiducial pseudorapidity range. The data are compared to fitted contributions described in the legend. The fit residuals normalised to the data uncertainty are shown at the bottom of each distribution.}
\label{fig:tff0}
\end{center}
\end{figure}

\section{Cross-section measurement}\label{sec:Measurement}

The inclusive \W boson production cross-section is measured in eight bins of muon $\eta$ between 2.0 and 4.5, and with the requirement that the muon has a \pt above 20\gevc. The cross-section in each pseudorapidity bin is defined as

\begin{equation*}
\cswpm(\eta_{i}) = \frac{1}{\lumi} \cdot \frac{\nc \cdot \rho^{\pm}(\eta_{i})}{\acc(\eta_{i}) \cdot \effrec(\eta_{i}) \cdot \effsel(\eta_{i})} \cdot \frac{1}{1 - \ffsr(\eta_{i})},
\end{equation*}

\noindent where \lumi is the integrated luminosity corresponding to the data set used in the analysis and $\nc$ is the total number of selected \wmn candidates. The signal purity, $\rho^{\pm}$, the acceptance, \acc, the reconstruction and the selection efficiencies, \effrec and \effsel, and the correction for \fsr, \ffsr, are determined for each $\eta$ bin. The cross-section in the range $2.0 < \eta< 4.5$ is obtained by summing over all bins.

\subsection{Acceptance}

The acceptance factor, \acc, is used to correct for the reduced \pt range of the fit. The correction is taken from the \wmn \resbos simulation and is defined as the fraction of generated events fulfilling the kinematic requirements of the measurement that have a muon transverse momentum smaller than 70\gevc. The average acceptance is 99.3\% and 99.1\% for \Wp and \Wm, respectively. In addition, no significant migration between $\eta$ bins is observed and no correction is applied.

\subsection{Reconstruction efficiencies}

The efficiency to reconstruct the muon produced in a \wmn decay, \effrec, is determined using data-driven techniques following Refs.~\cite{LHCb-PAPER-2012-008} and \cite{LHCb-PAPER-2012-036}. The reconstruction efficiency is factorised as the product of the tracking, the particle identification, the trigger and the global event cut efficiencies. The \zmm decay mode provides a highly pure sample of events that is used to determine efficiencies for high-\pt muons via a tag-and-probe method. The tag is a reconstructed track identified as a muon that triggered the event, while a second reconstructible object is used as the probe. The tag and the probe are required to have \pt greater than 20\gevc, a pseudorapidity in the range $2.0 < \eta < 4.5$, to be associated to the same primary vertex and have a combined invariant mass between 60 and 120\gevcc. The efficiency is determined as the fraction of tag-and-probe candidates that fulfil a specific requirement on the probe.

The tracking efficiency is determined using a probe track reconstructed by combining hits in the muon stations and the large-area silicon-strip detector, which are not used in the track finding. In order to reduce the background contribution, the accepted \mmm range is restricted to $70-110\gevcc$, and the tag and the probe are required to have an azimuthal separation greater than 0.1\rad. The efficiency depends on the probe $\eta$ value and varies between 87.2\% and 97.3\%. Simulation is used to correct for possible biases in the method, which range between 0.4\% and 2.0\%, and the corresponding systematic uncertainty is combined in quadrature with the statistical precision of the determination. The total uncertainty on the tracking efficiency varies between 0.4\% and 1.0\%.

The efficiency for identifying a muon is measured using a reconstructed track as the probe lepton. The tag and the probe are required to be isolated ($\ptcone < 2\gevc$) and to have an azimuthal separation greater than 2.7\rad to ensure a sample of very high purity. The use of the isolation requirement is investigated in simulation and proved not to bias the efficiency determination. The muon identification efficiency is found to vary between 95.6\% and 98.8\% as a function of the probe pseudorapidity. The corresponding uncertainty is about 0.2\% and accounts for the small amount of residual background.

The trigger efficiency is evaluated by using an identified muon as the probe object. To reduce the background to a negligible level, an isolation requirement ($\ptcone < 2\gevc$) is applied to the tag and the probe. The measurement is not biased by the requirement on the \ptcone, as verified using simulation. The efficiency depends on the $\eta$ value of the probe and ranges between 77.6\% and 81.5\%. The trigger efficiency uncertainty corresponds to the statistical uncertainty only and varies between 0.4\% and 0.6\%. To account for losses due to a high event occupancy in the hardware trigger, the global event cut efficiency is determined from data using a dimuon trigger with a relaxed threshold. The efficiency depends on the event multiplicity. After correcting for occupancy differences between \Z and \W events, this is determined to be $(95.9 \pm 1.1)\%$.

The reconstruction efficiencies are checked against simulation, for dependences on the muon charge as well as other quantities. No biases are observed.

\subsection{Selection efficiency}

The efficiency to select \wmn events, \effsel, defined as the fraction of events that fulfil the requirements used to identify the candidates, is measured using the \pw sample. Good agreement is observed between \pw data and \wmn simulation; however, the larger average \pt of \pw events must be accounted for. Simulation is used to asses the difference due to measuring the selection efficiency with muons produced in \Z instead of \W boson decays, which on average amounts to about 2\%. The statistical uncertainty of the data-driven determination and of the simulation-based correction are summed in quadrature. The efficiency is found to vary as a function of the muon pseudorapidity. No dependence on the lepton charge is observed except for the most forward $\eta$ bin. The measured \effsel ranges between 61.9\% and 70.0\% in the central bins with an uncertainty of about 0.5\%, but decreases to 49.0\% in the first pseudorapidity bin. The efficiency to select \Wp and \Wm in the range $4.0 < \eta < 4.5$ is $(34.1 \pm 0.8)\%$ and $(29.3 \pm 0.7)\%$, respectively.

\subsection{Final state radiation}

The \fsr correction, \ffsr, is evaluated using \photos interfaced to \pythia as the ratio of the number of generated events that after photon radiation fail the kinematic requirements to the number of generated decays that before \fsr satisfy the same criteria. The correction in bins of muon pseudorapidity is reported in Appendix~\ref{app:tables} (Table \ref{tab:fsr}).

\subsection{Systematic uncertainties}

The contributions to the systematic uncertainty considered in the analysis are the shape and the normalisation of the templates used in the fit, the reconstruction and the selection efficiencies, the acceptance and the \fsr corrections, and the luminosity determination. A summary of the systematic uncertainties on the \Wp and \Wm cross-sections and their ratio is given in Table \ref{tab:systematic}.

\begin{table}[!t]
\begin{center}
\caption{Summary of the systematic uncertainties on the inclusive cross-sections and their ratio.}
\begin{tabular}{lccc}
Source & $\Delta\cswp$ [\%] & $\Delta\cswm$ [\%] & $\Delta\ratio$ [\%] \\ \hline
Template shape & 0.28 & 0.39 & 0.59 \\
Template normalisation & 0.10 & 0.10 & 0.06 \\
Reconstruction efficiency & 1.21 & 1.20 & 0.12 \\
Selection efficiency & 0.33 & 0.32 & 0.18 \\
Acceptance and FSR & 0.18 & 0.12 & 0.21 \\
Luminosity & 1.71 & 1.71 & --- \\
\end{tabular}
\label{tab:systematic}
\end{center}
\end{table}

An estimate of the uncertainty arising from the choice of the template shapes is evaluated by refitting the data with different \pt distributions for the signal and the main background components, and combining the observed variations in the results in quadrature.
The weights that correct the \wmn and \zmm templates to reproduce the \resbos muon \pt distributions are smeared by their statistical uncertainty. The observed change on the \Wp (\Wm) cross-section is 0.01\% (0.05)\%.
The shape of the decay-in-flight template is modified by constraining the relative normalisation between positively or negatively charged particles in each $\eta$ bin according to fractions observed in randomly triggered events. The \Wp and \Wm cross-sections vary by 0.23\% and 0.35\% respectively.
In addition, the correction applied to simulation-based templates that accounts for efficiency differences with respect to data is modified by its statistical uncertainty, inducing a variation of 0.16\% (0.15)\% in \cswp (\cswm). The larger template shape uncertainty assigned to the \Wm cross-section accounts for the better description by the fit model of the \pt spectrum of positive muons, as shown in Fig.~\ref{fig:tff0}.

Similarly, the normalisations of the constrained templates are varied independently and the largest deviations corresponding to each source are summed in quadrature.
The electroweak template normalisation is shifted by $\pm 1\,\sigma$, which corresponds to a change of the \Wp and \Wm production cross-sections of 0.10\%.
The residual amount of heavy flavour events is varied by its statistical uncertainty. No change is observed on \cswp and \cswm. Because the heavy flavour template shape is very similar to the decay-in-flight component, variations to the former are compensated by a change in the normalisation of the latter, with small effects on the results.

The reconstruction and the selection efficiencies are measured from data and the uncertainty of each determination is taken as an estimate of the corresponding systematic uncertainty. The resulting uncertainty on the \W cross-sections is 1.2\% for \effrec, and 0.3\% for \effsel.

The acceptance and \fsr corrections are evaluated using simulation. The statistical uncertainty of \acc and the total uncertainty of \ffsr, which includes extra sources of theoretical uncertainties, are summed in quadrature and result in a 0.18\% (0.12)\% systematic uncertainty on the \Wp (\Wm) boson cross-section.

The luminosity determination has a precision of 1.71\%~\cite{LHCB-PAPER-2014-047} and is quoted separately from the other sources of systematic uncertainty.

To check for possible detector-induced asymmetries, data samples collected with opposite magnet polarity are analysed separately. No significant discrepancy is observed on the \Wp to \Wm cross-section ratio, which does not depend on the luminosity determination and is less sensitive to efficiency biases, and no systematic uncertainty is added.

\section{Results}\label{sec:Results}

The \wpmn and \wmmn production cross-sections for muons with a \pt exceeding 20\gevc and $2.0 < \eta < 4.5$ are measured to be
\begin{align*}
\cswp &= \csp, \\
\cswm &= \csm,
\end{align*}

\noindent where the first uncertainty is statistical, the second is systematic and the third is due to the luminosity determination. Aside from the luminosity contribution, measurements are dominated by the limited knowledge of the reconstruction efficiency. The correlation coefficient between the \Wp and \Wm cross-sections for the statistical and systematic uncertainties combined corresponds to 0.83, which increases to 0.94 when including the luminosity uncertainty. The full correlation matrix in bins of muon pseudorapidity is reported in Appendix~\ref{app:tables} (Table \ref{tab:rho}).

The ratio of the \Wp and the \Wm cross-sections is determined to be

\begin{equation*}
\ratio = \frac{\cswp}{\cswm} = \csr,
\end{equation*}

\noindent where the uncertainties are statistical and systematic. The luminosity uncertainty cancels, while the systematic uncertainties associated with the efficiencies are reduced due to correlations.

The muon charge asymmetry is measured in each pseudorapidity bin according to

\begin{equation*}
\asy(\eta_{i}) = \frac{\cswp(\eta_{i}) - \cswm(\eta_{i})}{\cswp(\eta_{i}) + \cswm(\eta_{i})}.
\end{equation*}

\begin{figure}[!t]
\begin{center}
\includegraphics[width=.7\textwidth]{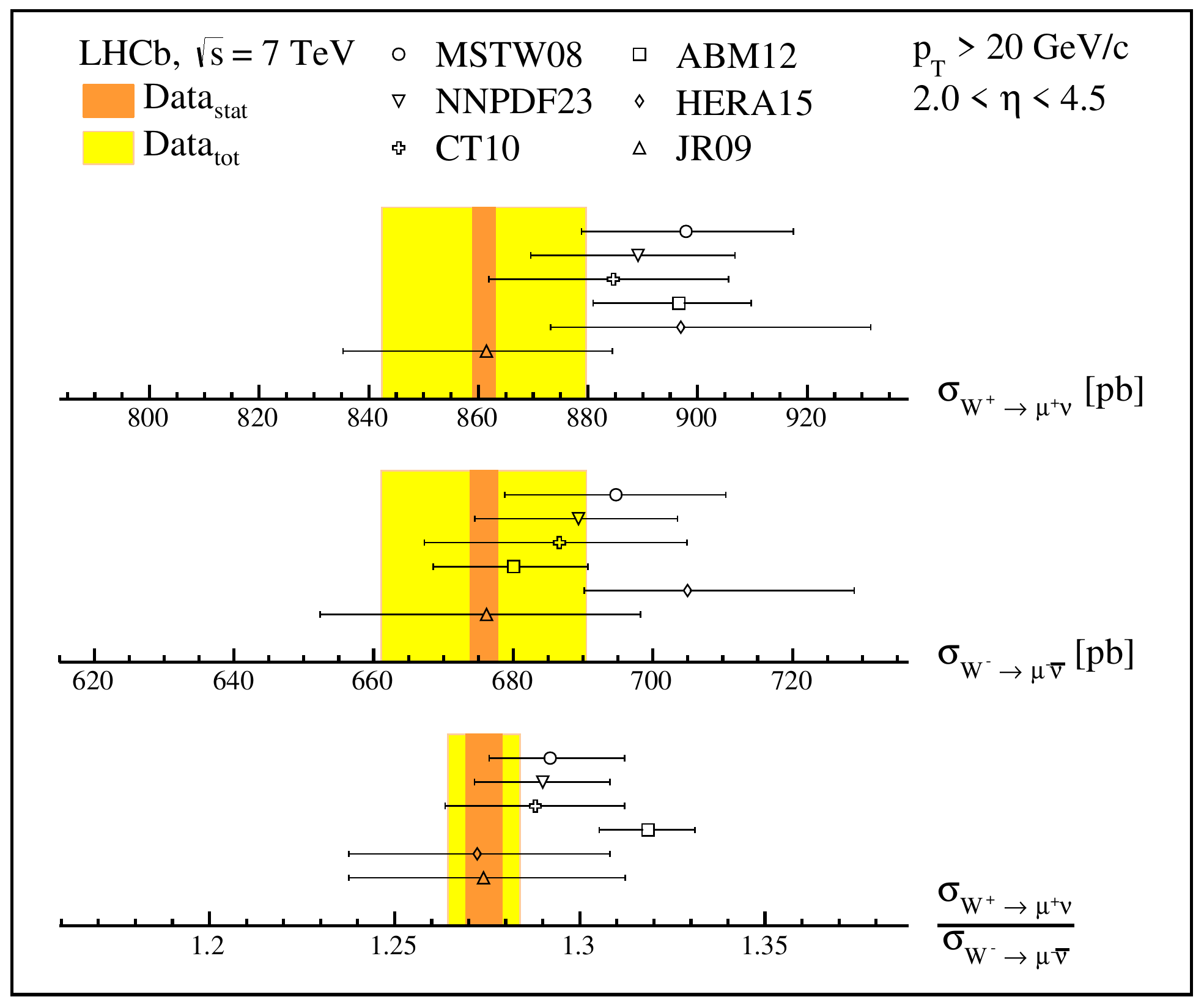}
\caption{Summary of the \W cross-section determinations. Measurements, represented as bands corresponding to the statistical (orange) and total (yellow) uncertainty, are compared to \nnlo predictions for various parameterisations of the \pdfs (black markers).}
\label{fig:summary}
\end{center}
\end{figure}

\begin{figure}[!t]
\begin{center}
\includegraphics[width=.7\textwidth]{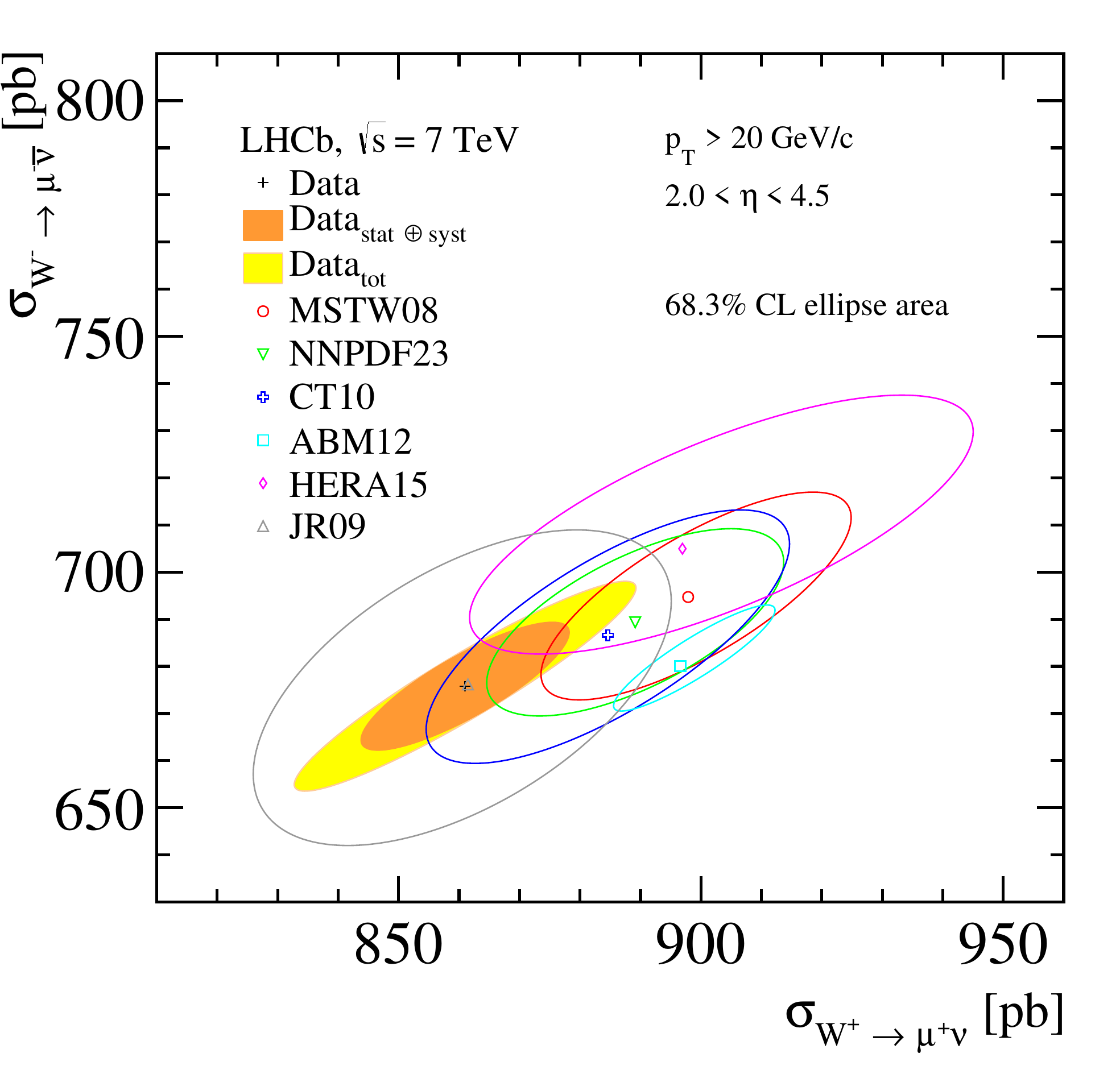}
\caption{Two-dimensional plot of the measured (total (yellow) and excluding the luminosity (orange) uncertainty) \Wp and \Wm cross-sections compared to \nnlo predictions for various parameterisations of the \pdfs (coloured markers). The uncertainty of the theoretical predictions corresponds to the \pdf uncertainty only; the correlation is determined using the different error eigenvector sets. The ellipses correspond to a $68.3\%$ CL coverage.}
\label{fig:ellipse}
\end{center}
\end{figure}

\begin{figure}[!t]
\begin{center}
\includegraphics[width=.7\textwidth]{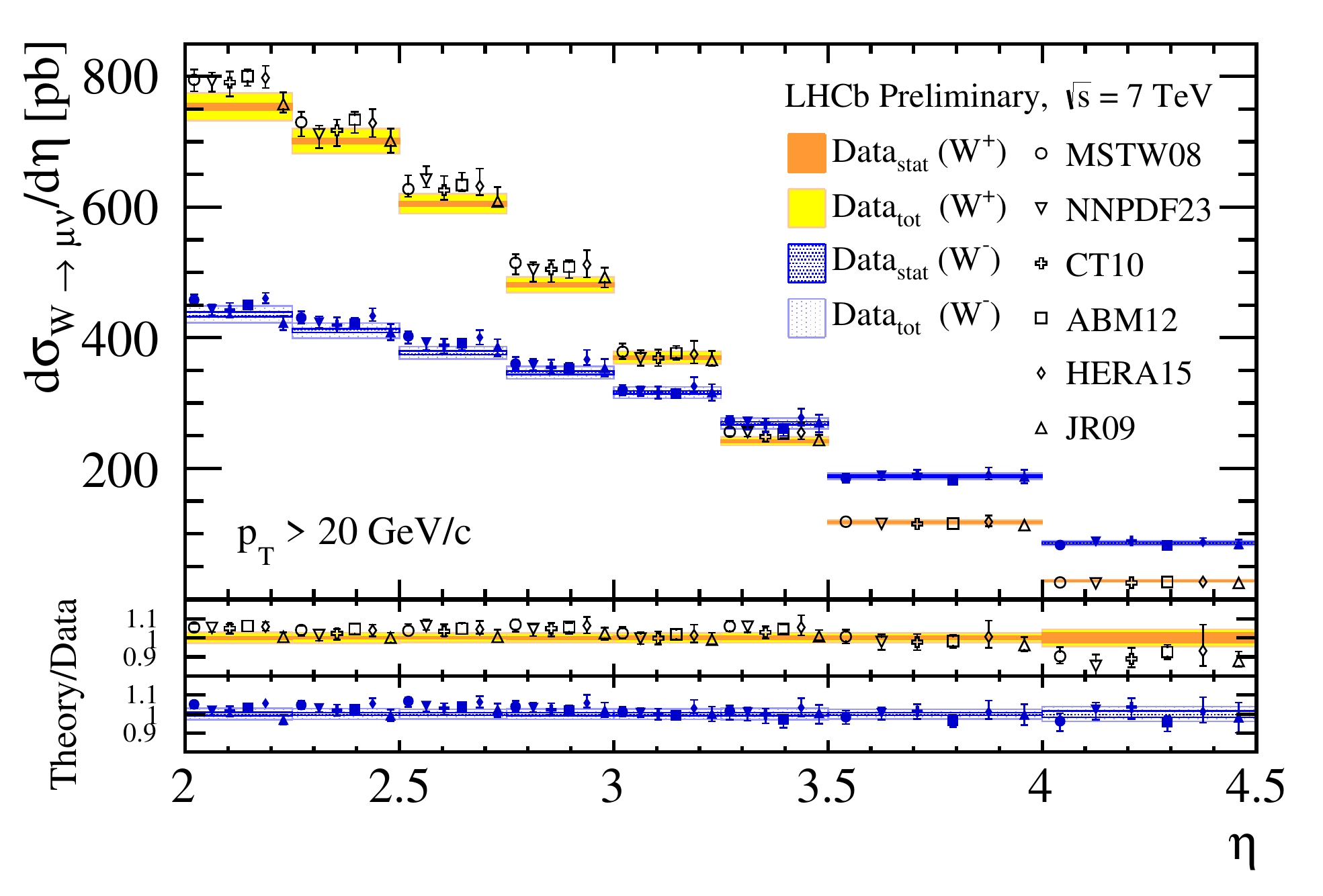}
\caption{Differential \Wp and \Wm cross-section in bins of muon pseudorapidity. Measurements, represented as bands corresponding to the statistical (orange (blue) for \Wp (\Wm)) and total (yellow (light blue) for \Wp(\Wm)) uncertainty, are compared to \nnlo predictions with different parameterisations of the \pdfs (black (blue) markers for \Wp (\Wm), displaced horizontally for presentation).}
\label{fig:cs}
\end{center}

\begin{center}
\includegraphics[width=.7\textwidth]{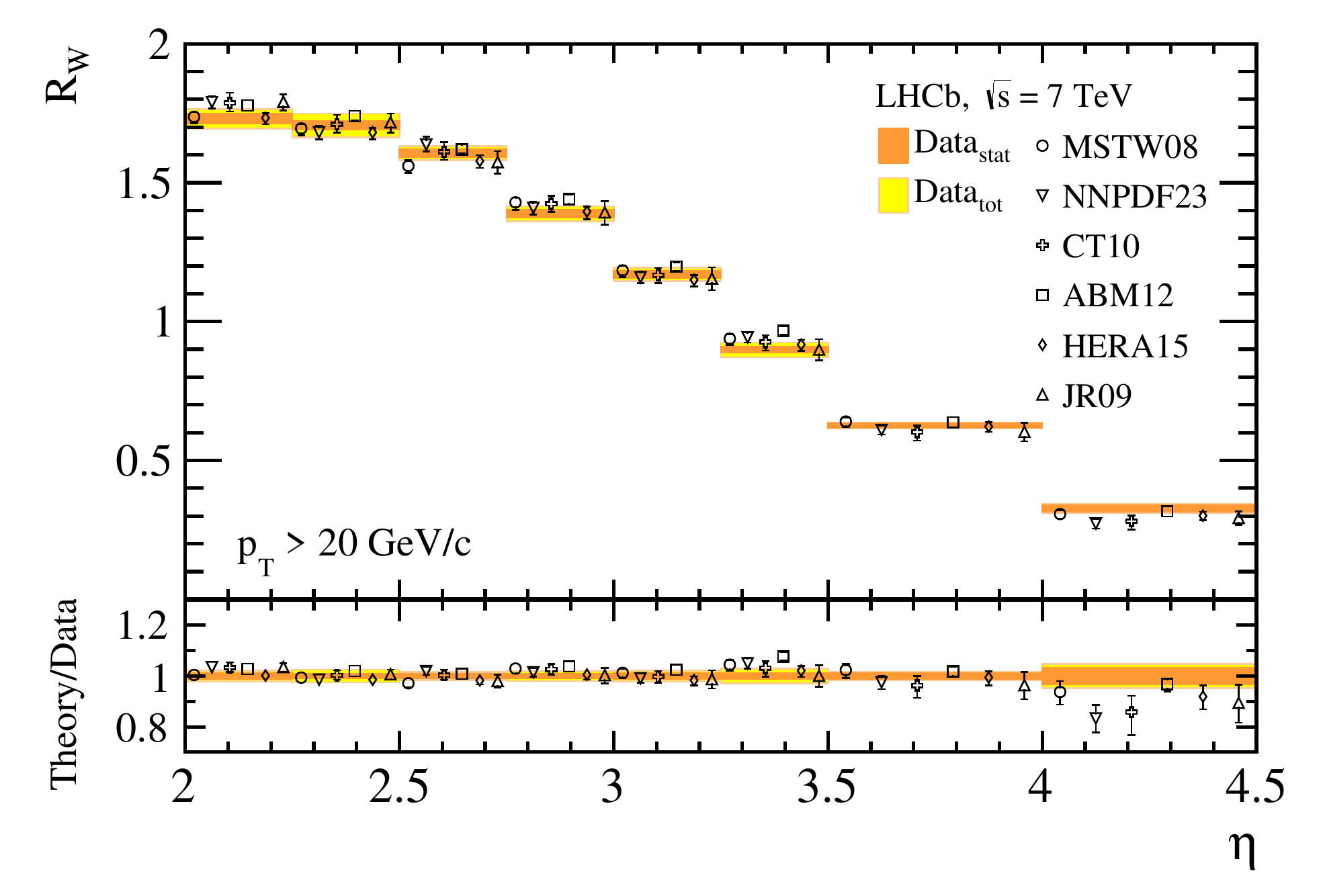}
\caption{Ratio of \Wp to \Wm cross-sections in bins of muon pseudorapidity. Measurements, represented as bands corresponding to the statistical (orange) and total (yellow) uncertainty, are compared to \nnlo predictions for various parameterisations of the \pdfs (black markers, displaced horizontally for presentation).}
\label{fig:csr}
\end{center}
\end{figure}

\begin{figure}[!t]
\begin{center}
\includegraphics[width=.7\textwidth]{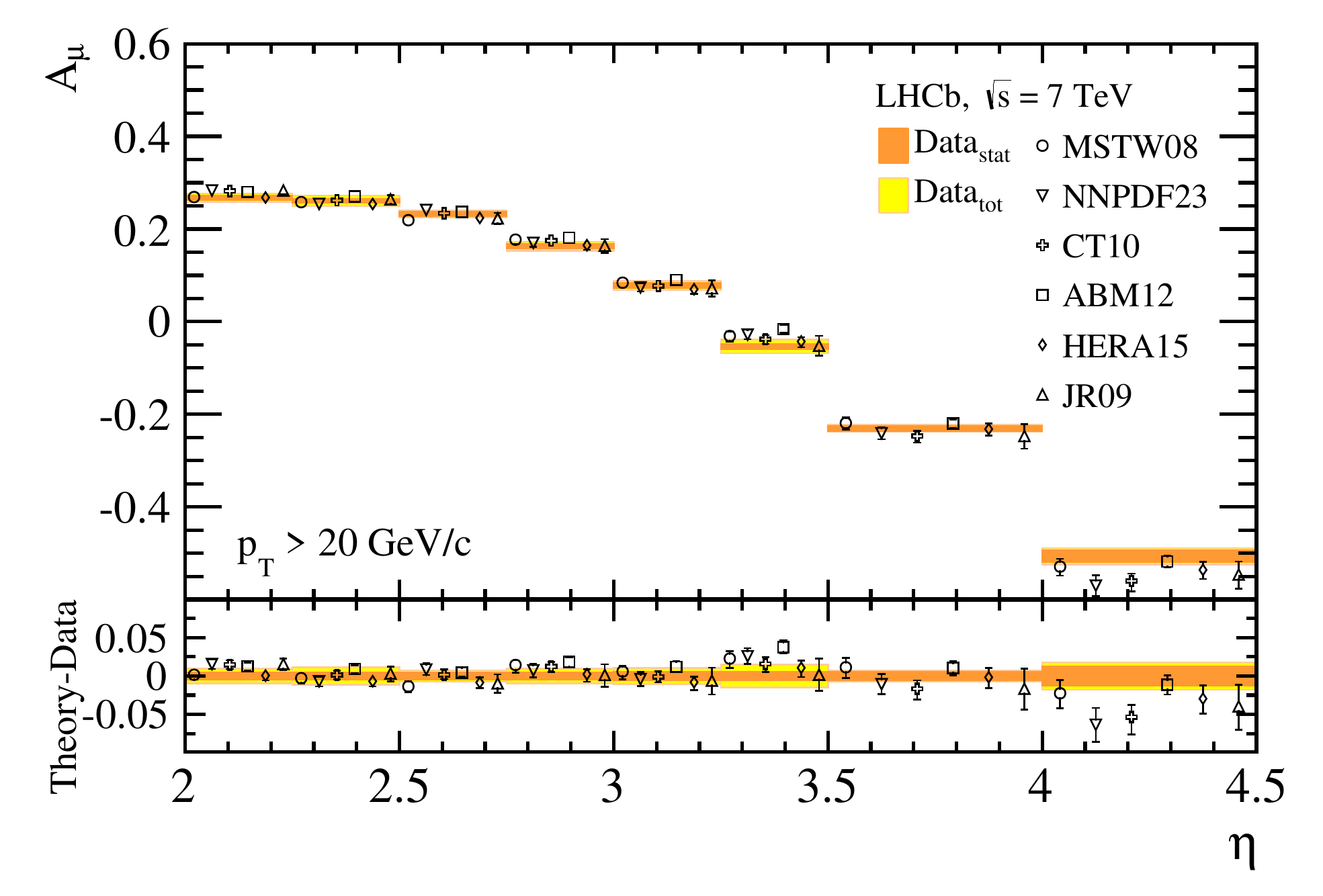}
\caption{Lepton charge asymmetry in bins of muon pseudorapidity. Measurements, represented as bands corresponding to the statistical (orange) and total (yellow) uncertainty, are compared to \nnlo predictions for various parameterisations of the \pdfs (black markers, displaced horizontally for presentation).}
\label{fig:csa}
\end{center}
\end{figure}

A summary of the measurements for muons with $\pt > 20\gevc$ and $2.0 < \eta < 4.5$ is presented in Fig.~\ref{fig:summary}. The cross-sections for inclusive \Wp and \Wm production in a two-dimensional plot are shown in Fig.~\ref{fig:ellipse}, where the ellipses correspond to 68.3\% CL coverage. The differential \Wp and \Wm cross-sections and the \Wp to \Wm cross-section ratio as a function of the muon pseudorapidity are shown in Figs.~\ref{fig:cs} and \ref{fig:csr}, while the lepton charge asymmetry is presented in Fig.~\ref{fig:csa}.  Measurements are compared to predictions at \nnlo in \qcd with different parameterisations of the \pdfs. Results are generally in good agreement with theoretical calculations. Cross-section and ratio determinations in bins of muon $\eta$ are tabulated in Appendix~\ref{app:tables} (Tables~\ref{tab:cs}, \ref{tab:csr} and \ref{tab:csa}).

\section{Conclusions} \label{sec:Conclusions}

Measurements of inclusive \wmn production in \pp collisions at a centre-of-mass energy of \sqs = 7\tev using a data set corresponding to an integrated luminosity of \dataset recorded by the \lhcb experiment are presented. The cross-section for \Wp and \Wm boson production, as well as their ratio, \ratio, and charge asymmetry, \asy, are measured for muons with $\pt > 20\gevc$ and $2.0 < \eta < 4.5$. Results are compared to Standard Model predictions calculated at \nnlo in perturbative \qcd, which agree well with the data. The cross-section measurements presented in this paper provide constraints on the determinations of the proton \pdfs, with a total uncertainty that is comparable with, or in some cases better, than the uncertainty on the theory calculations. The precision on the \Wp to \Wm cross-section ratio, which is improved by a factor of about two with respect to the previous result~\cite{LHCb-PAPER-2012-008}, allows the Standard Model to be tested with a sub-percent accuracy.

\section*{Acknowledgements}

\noindent We express our gratitude to our colleagues in the CERN accelerator departments for the excellent performance of the LHC. We thank the technical and administrative staff at the LHCb institutes. We acknowledge support from CERN and from the national agencies: CAPES, CNPq, FAPERJ and FINEP (Brazil); NSFC (China); CNRS/IN2P3 (France); BMBF, DFG, HGF and MPG (Germany); SFI (Ireland); INFN (Italy); FOM and NWO (The Netherlands); MNiSW and NCN (Poland); MEN/IFA (Romania); MinES and FANO (Russia); MinECo (Spain); SNSF and SER (Switzerland); NASU (Ukraine); STFC (United Kingdom); NSF (USA). The Tier1 computing centres are supported by IN2P3 (France), KIT and BMBF (Germany), INFN (Italy), NWO and SURF (The Netherlands), PIC (Spain), GridPP (United Kingdom). We are indebted to the communities behind the multiple open source software packages on which we depend. We are also thankful for the computing resources and the access to software R\&D tools provided by Yandex LLC (Russia). Individual groups or members have received support from EPLANET, Marie Sk\l{}odowska-Curie Actions and ERC (European Union), Conseil g\'{e}n\'{e}ral de Haute-Savoie, Labex ENIGMASS and OCEVU, R\'{e}gion Auvergne (France), RFBR (Russia), XuntaGal and GENCAT (Spain), Royal Society and Royal Commission for the Exhibition of 1851 (United Kingdom).

\clearpage

{\noindent\bf\Large Appendices}

\appendix

\section{Fit}\label{app:fit}

\begin{figure}[!h]
\begin{center}
\includegraphics[width=.49\textwidth]{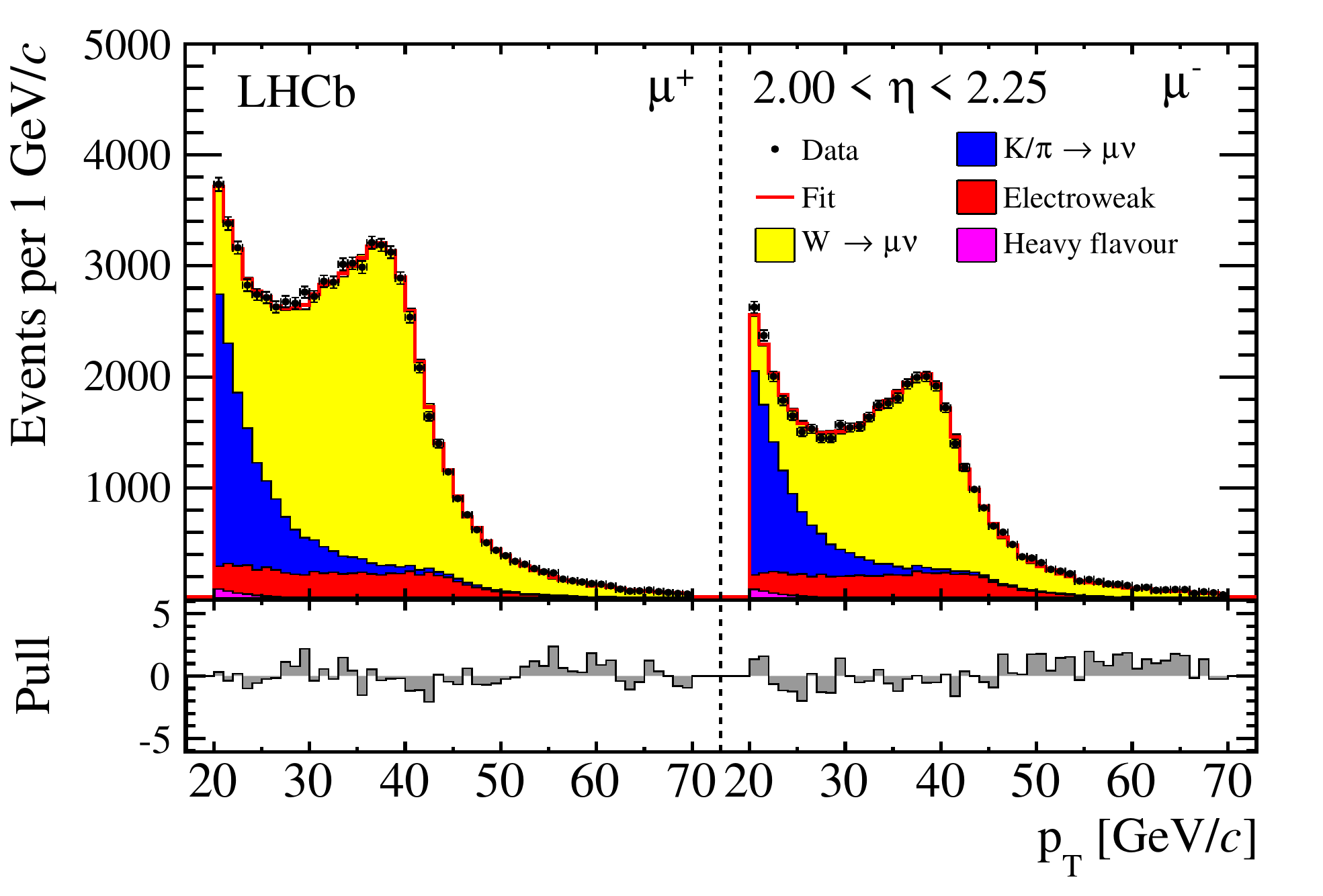}
\includegraphics[width=.49\textwidth]{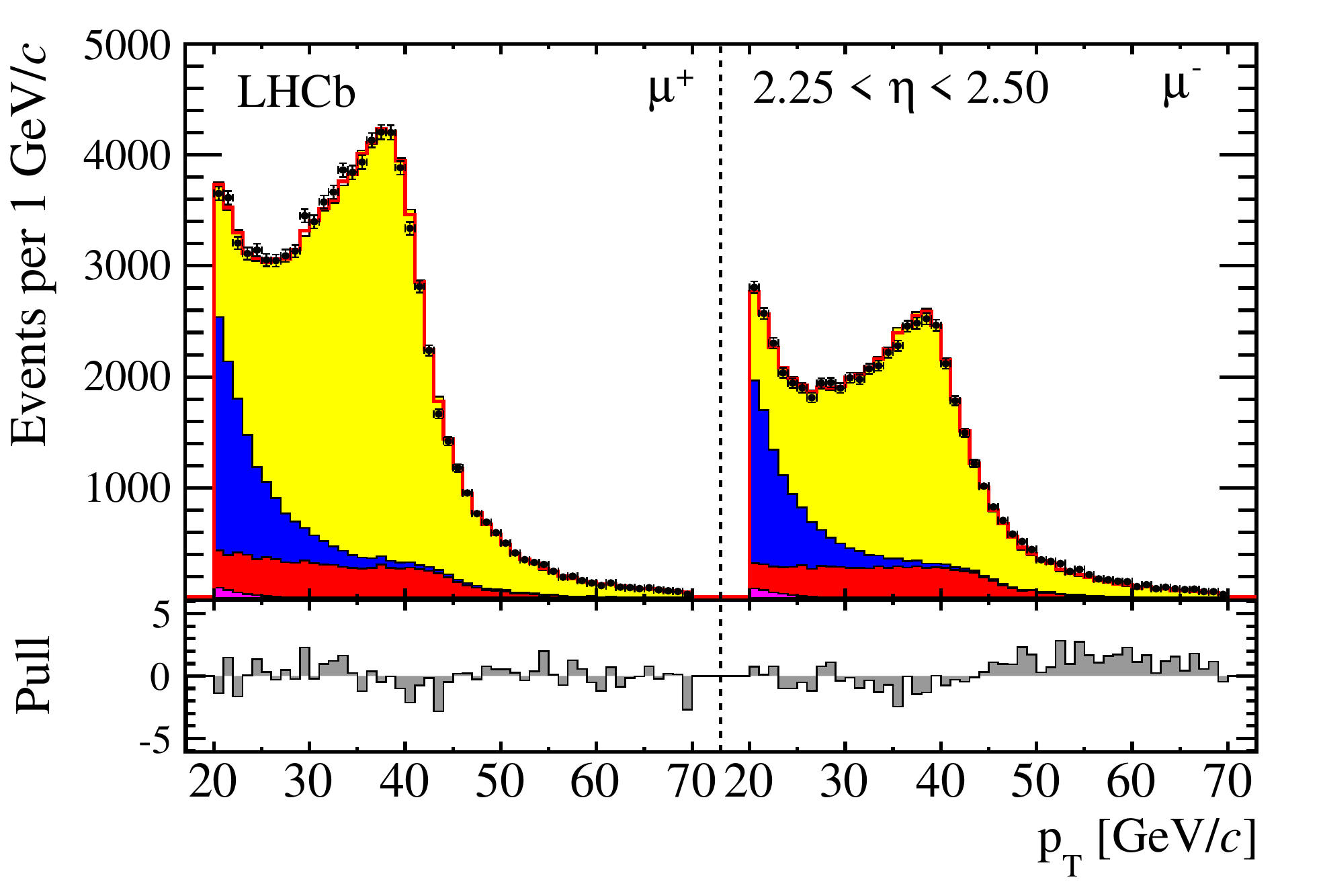}
\includegraphics[width=.49\textwidth]{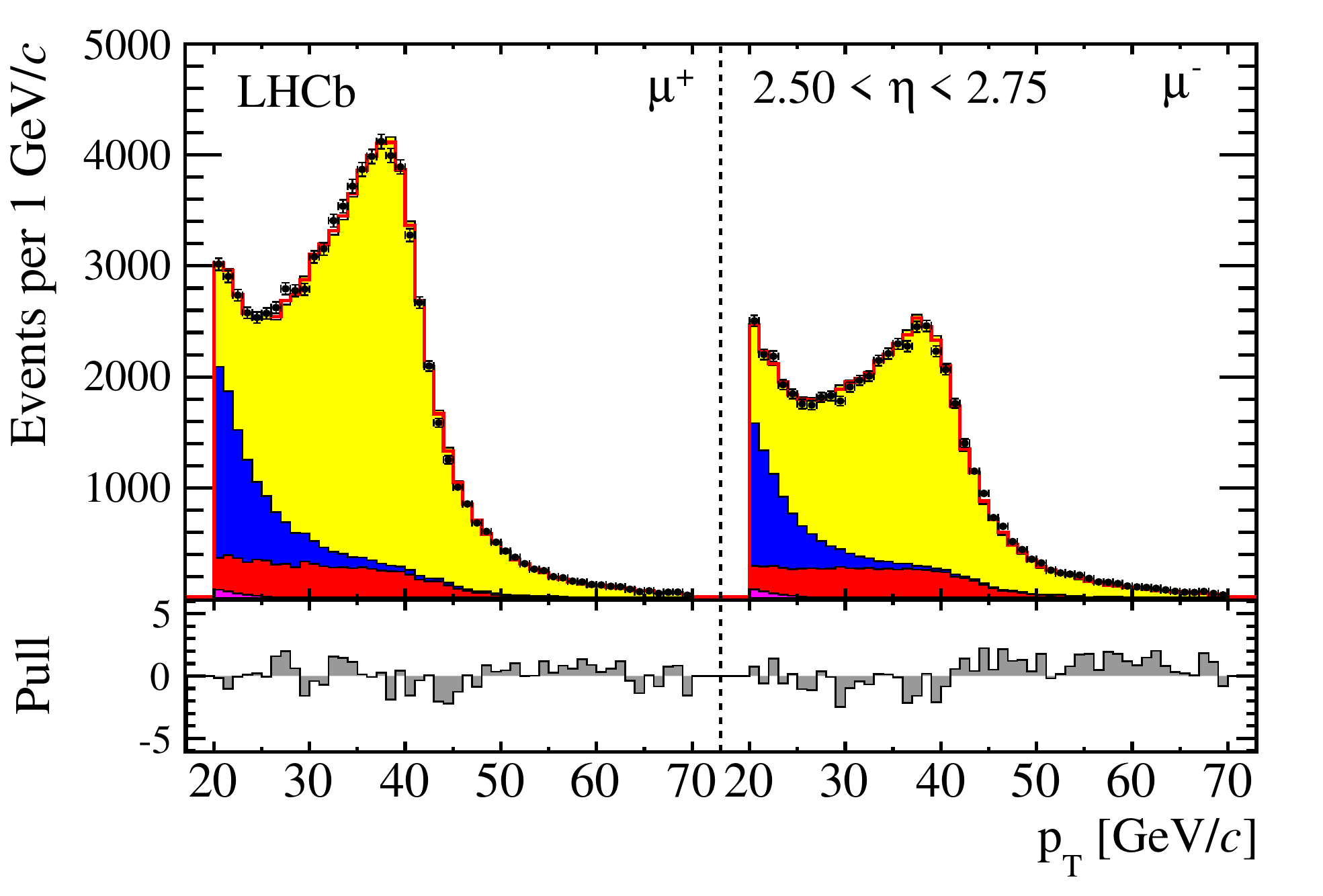}
\includegraphics[width=.49\textwidth]{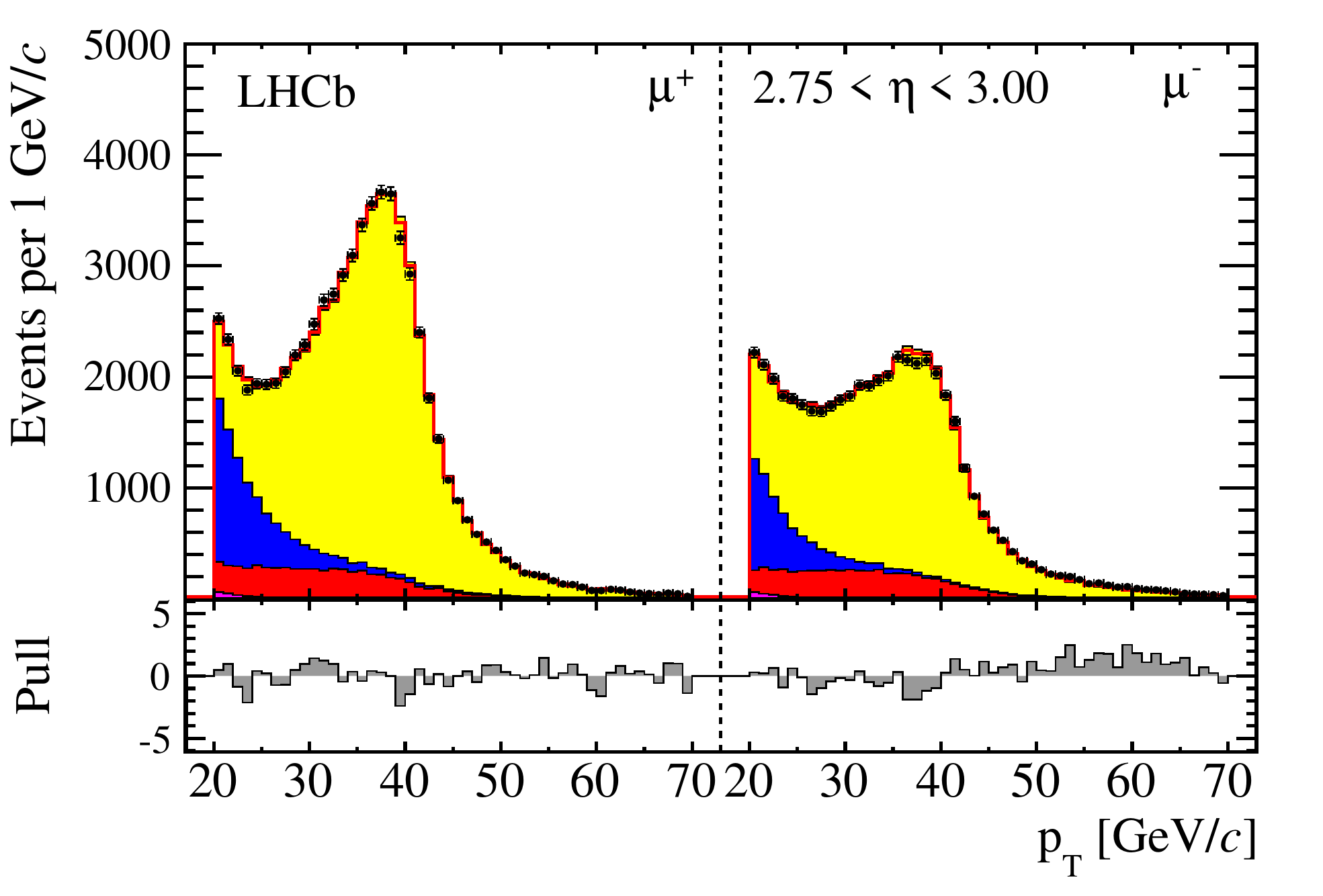}
\caption{Transverse momentum distribution of the (left panel) positive and (right panel) negative muon candidates in eight bins of pseudorapidity. The data are compared to fitted contributions described in the legend. The fit residuals normalised to the data uncertainty are shown at the bottom of each distribution.}
\label{fig:tff1}
\end{center}
\end{figure}

\begin{figure}[!h]
\begin{center}
\includegraphics[width=.49\textwidth]{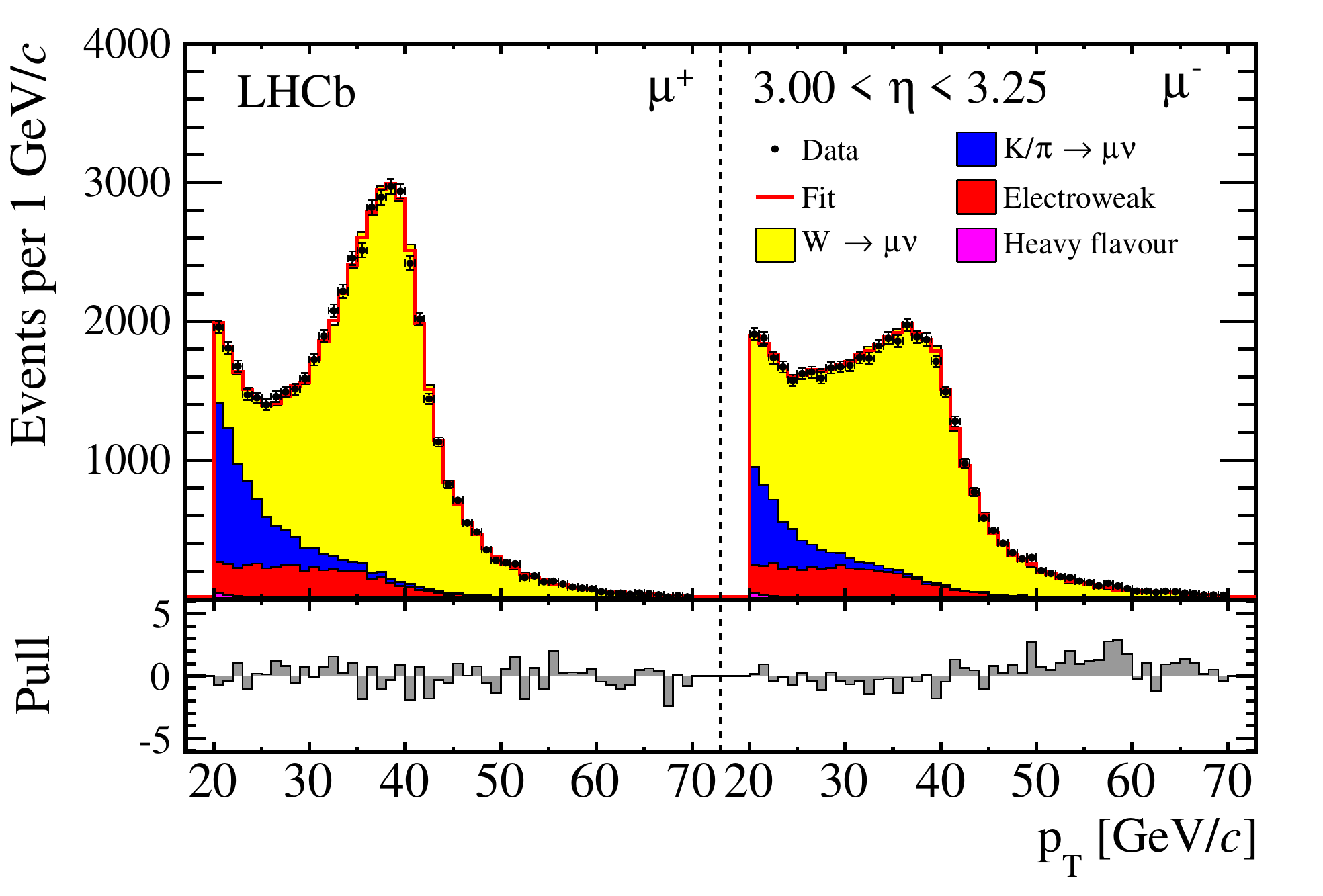}
\includegraphics[width=.49\textwidth]{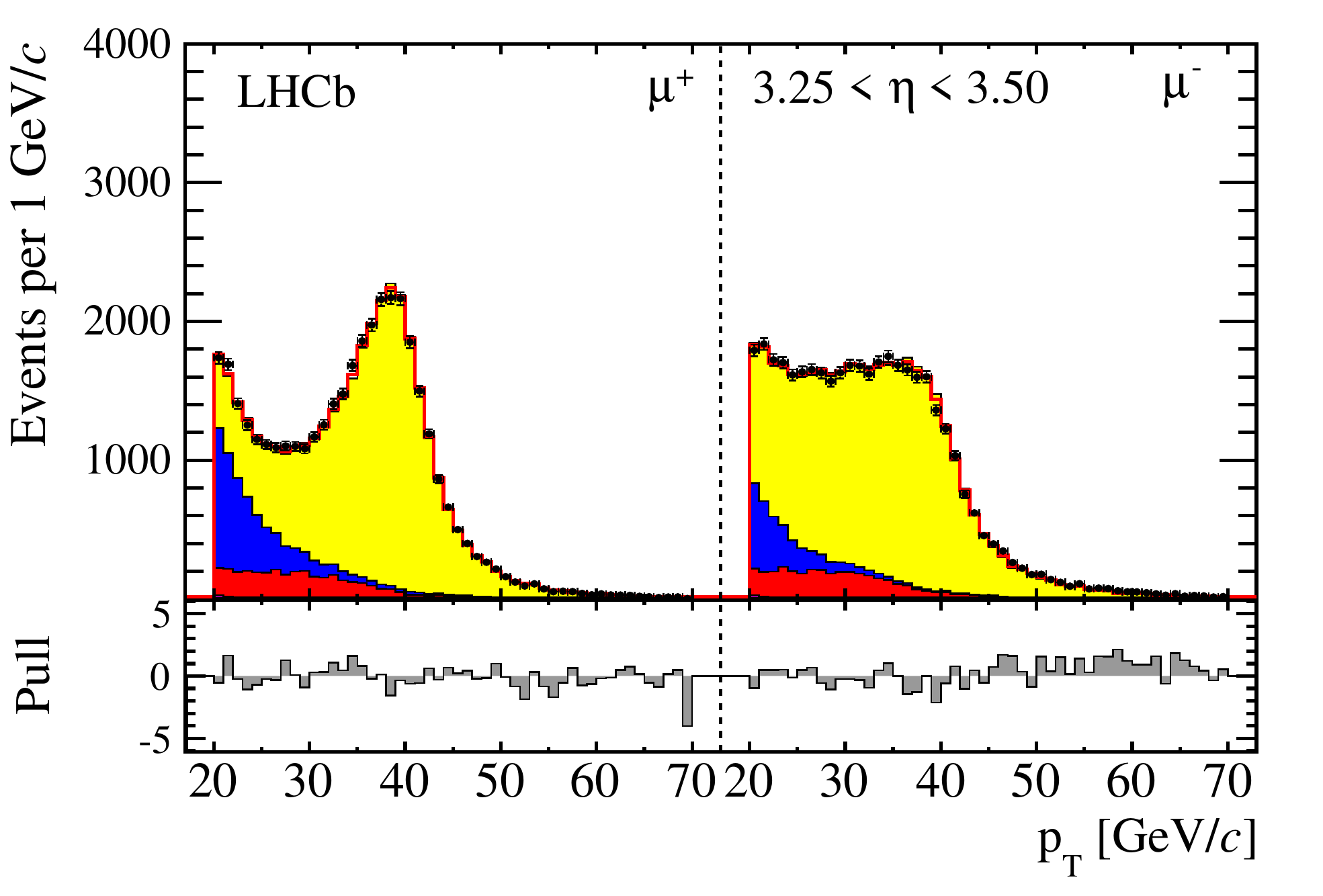}
\includegraphics[width=.49\textwidth]{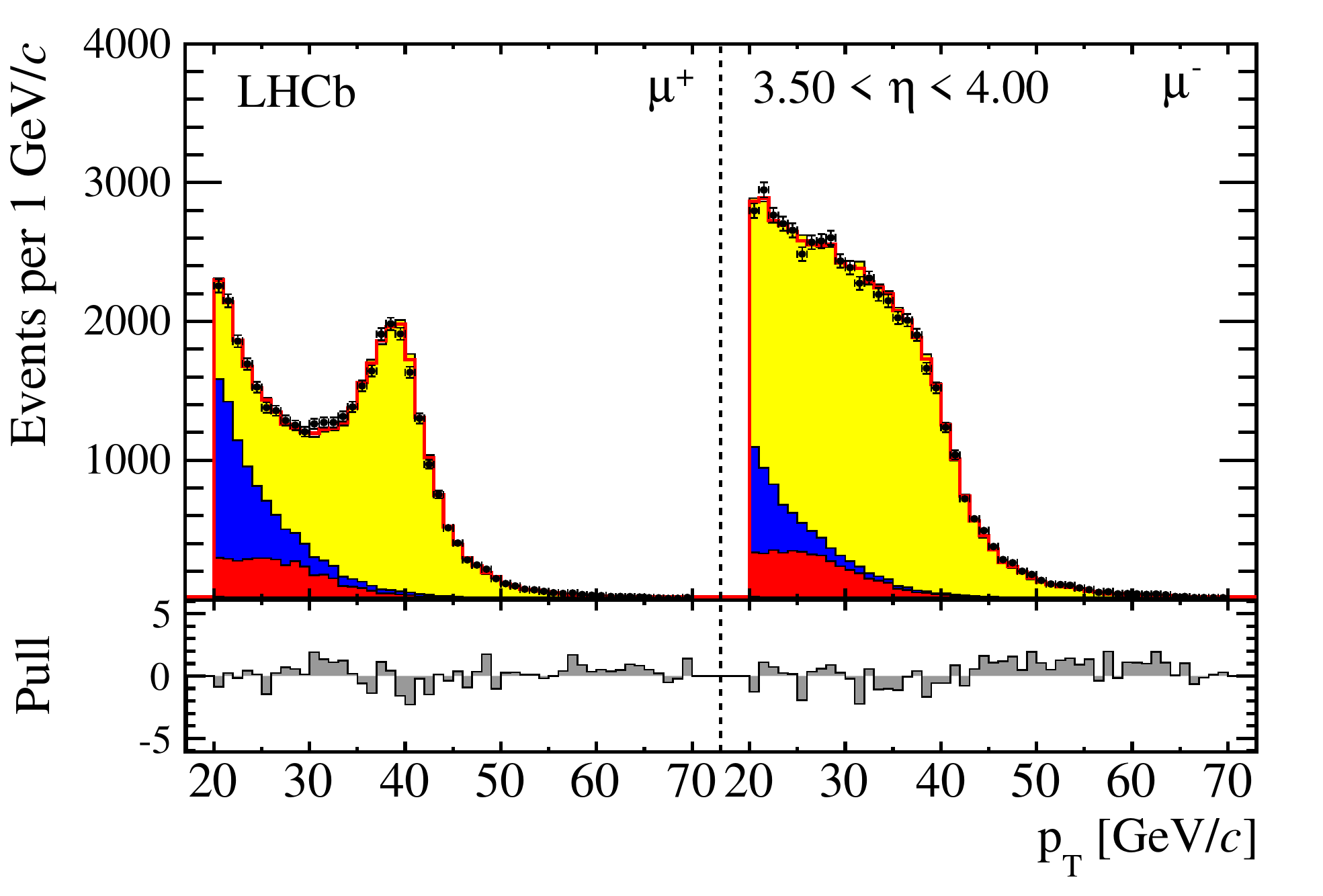}
\includegraphics[width=.49\textwidth]{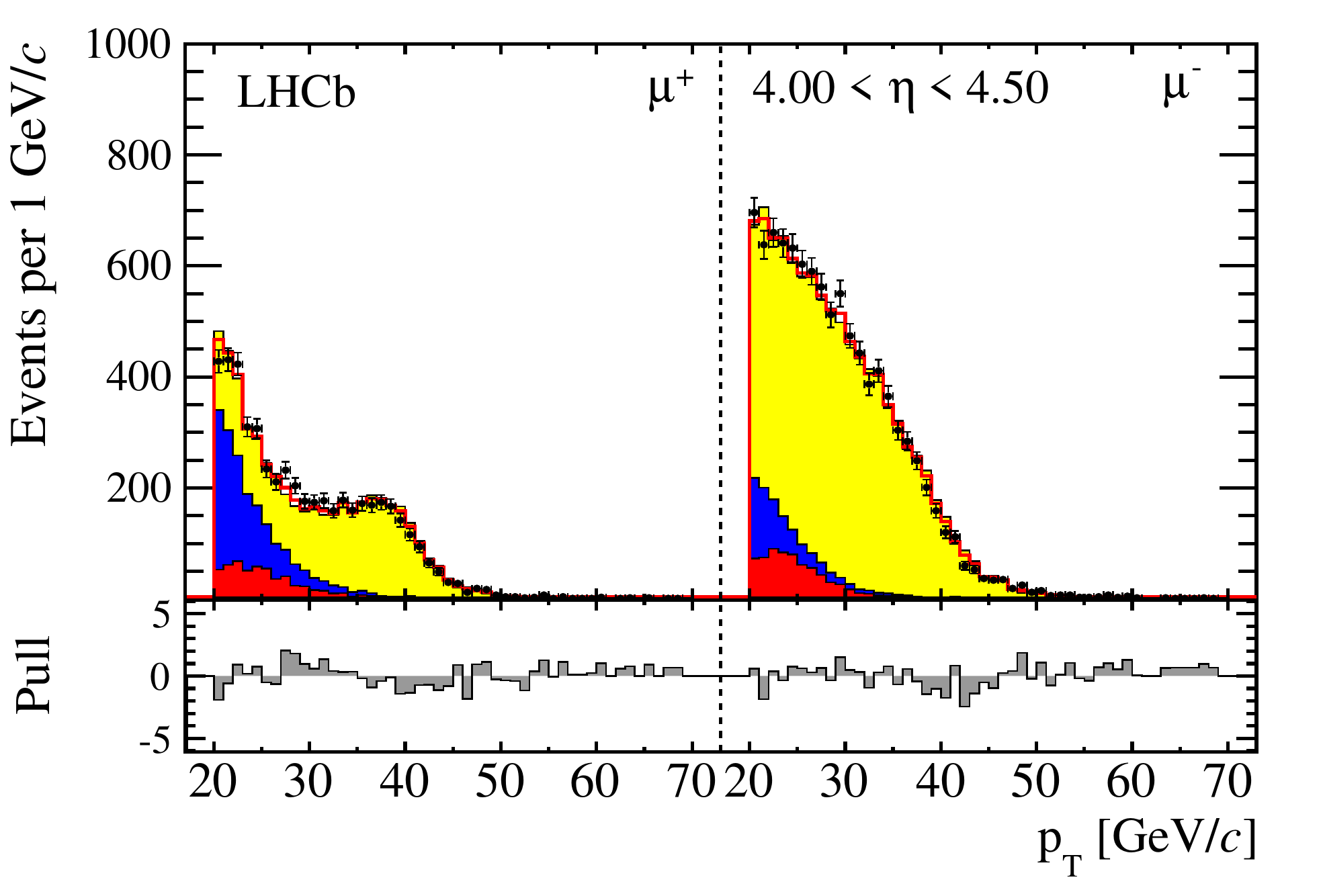}
\caption{Transverse momentum distribution of the (left panel) positive and (right panel) negative muon candidates in eight bins of pseudorapidity. The data are compared to fitted contributions described in the legend. The fit residuals normalised to the data uncertainty are shown at the bottom of each distribution.}
\label{fig:tff2}
\end{center}
\end{figure}

\clearpage

\section{Tables}\label{app:tables}

\begin{table}[!h]
\caption{Correction for final state radiation for \wpmn and \wmmn in bins of muon pseudorapidity. No loss due \fsr is observed for \Wm in the last pseudorapidity bin because of lack of statistics.}
\label{tab:fsr}
\begin{center}
\begin{tabular}{ccc} \hline
$\eta$ & $f^{+}_{\textrm{\fsr}}$ [\%] & $f^{-}_{\textrm{\fsr}}$ [\%] \\ \hline
$(2.00, 2.25)$ & $1.62 \pm 0.39$ & $1.84 \pm 0.30$ \\
$(2.25, 2.50)$ & $1.75 \pm 0.41$ & $1.46 \pm 0.28$ \\
$(2.50, 2.75)$ & $2.41 \pm 0.50$ & $1.03 \pm 0.25$ \\
$(2.75, 3.00)$ & $1.43 \pm 0.40$ & $0.72 \pm 0.23$ \\
$(3.00, 3.25)$ & $2.01 \pm 0.50$ & $0.93 \pm 0.31$ \\
$(3.25, 3.50)$ & $1.51 \pm 0.48$ & $1.71 \pm 0.51$ \\
$(3.50, 4.00)$ & $2.32 \pm 0.50$ & $1.22 \pm 0.45$ \\
$(4.00, 4.50)$ & $2.05 \pm 0.47$ & 0 \\ \hline
\end{tabular}
\end{center}

\vspace{1cm}

\caption{Inclusive cross-section for \Wp and \Wm boson production in bins of muon pseudorapidity. The uncertainties are statistical, systematic and luminosity.}
\label{tab:cs}
\begin{center}
\begin{tabular}{rrr} \hline
\multicolumn{1}{c}{$\eta$} & \multicolumn{1}{c}{\cswp [\pb]} & \multicolumn{1}{c}{\cswm [\pb]} \\ \hline
$(2.00, 2.25)$ & $188.4 \pm 1.2 \pm 4.0 \pm 3.2$ & $108.9 \pm 0.9 \pm 2.4 \pm 1.9$ \\
$(2.25, 2.50)$ & $175.3 \pm 0.9 \pm 3.6 \pm 3.0$ & $102.8 \pm 0.7 \pm 2.2 \pm 1.8$ \\
$(2.50, 2.75)$ & $151.3 \pm 0.8 \pm 2.6 \pm 2.6$ & $94.2 \pm 0.7 \pm 1.6 \pm 1.6$ \\
$(2.75, 3.00)$ & $120.3 \pm 0.7 \pm 2.0 \pm 2.1$ & $86.7 \pm 0.6 \pm 1.8 \pm 1.5$ \\
$(3.00, 3.25)$ & $92.4 \pm 0.6 \pm 1.6 \pm 1.6$ & $79.0 \pm 0.6 \pm 1.6 \pm 1.4$ \\
$(3.25, 3.50)$ & $60.4 \pm 0.5 \pm 1.1 \pm 1.0$ & $67.2 \pm 0.5 \pm 1.6 \pm 1.1$ \\
$(3.50, 4.00)$ & $58.8 \pm 0.5 \pm 1.0 \pm 1.0$ & $94.0 \pm 0.7 \pm 1.6 \pm 1.6$ \\
$(4.00, 4.50)$ & $14.1 \pm 0.4 \pm 0.4 \pm 0.2$ & $42.9 \pm 0.7 \pm 1.3 \pm 0.7$ \\ \hline
\end{tabular}
\end{center}
\end{table}

\begin{table}[!h]
\caption{Ratio of \Wp to \Wm cross-section in bins of muon pseudorapidity. The uncertainties are statistical and systematic.}
\label{tab:csr}
\begin{center}
\begin{tabular}{rr} \hline
\multicolumn{1}{c}{$\eta$} & \multicolumn{1}{c}{\ratio} \\ \hline
$(2.00, 2.25)$ & $1.730 \pm 0.018 \pm 0.030$ \\
$(2.25, 2.50)$ & $1.706 \pm 0.015 \pm 0.040$ \\
$(2.50, 2.75)$ & $1.606 \pm 0.014 \pm 0.021$ \\
$(2.75, 3.00)$ & $1.388 \pm 0.013 \pm 0.024$ \\
$(3.00, 3.25)$ & $1.169 \pm 0.012 \pm 0.021$ \\
$(3.25, 3.50)$ & $0.898 \pm 0.010 \pm 0.025$ \\
$(3.50, 4.00)$ & $0.626 \pm 0.007 \pm 0.006$ \\
$(4.00, 4.50)$ & $0.328 \pm 0.011 \pm 0.011$ \\ \hline
\end{tabular}
\end{center}

\vspace{1cm}

\caption{Lepton charge asymmetry in bins of muon pseudorapidity. The uncertainties are statistical and systematic.}
\label{tab:csa}
\begin{center}
\begin{tabular}{rr} \hline
\multicolumn{1}{c}{$\eta$} & \multicolumn{1}{c}{\asy [\%]} \\ \hline
$(2.00, 2.25)$ & $26.74 \pm 0.48 \pm 0.82$ \\
$(2.25, 2.50)$ & $26.08 \pm 0.41 \pm 1.09$ \\
$(2.50, 2.75)$ & $23.25 \pm 0.42 \pm 0.60$ \\
$(2.75, 3.00)$ & $16.26 \pm 0.46 \pm 0.84$ \\
$(3.00, 3.25)$ & $7.81 \pm 0.50 \pm 0.90$ \\
$(3.25, 3.50)$ & $-5.37 \pm 0.57 \pm 1.35$ \\
$(3.50, 4.00)$ & $-23.04 \pm 0.52 \pm 0.49$ \\
$(4.00, 4.50)$ & $-50.65 \pm 1.22 \pm 1.30$ \\ \hline
\end{tabular}
\end{center}
\end{table}

\begin{table}[!h]
\caption{Correlation coefficients (statistical and systematic uncertainties) between \Wp and \Wm cross-sections in bins of muon pseudorapidity. The luminosity uncertainty is not included.}
\label{tab:rho}
\begin{center}
\begin{scriptsize}
\setlength{\tabcolsep}{3pt}
\renewcommand{\arraystretch}{2.75}
\begin{tabular}{c|rr|rr|rr|rr|rr|rr|rr|rr|c}

& \multicolumn{2}{c|}{$(2.00, 2.25)$} & \multicolumn{2}{c|}{$(2.25, 2.50)$} & \multicolumn{2}{c|}{$(2.50, 2.75)$} & \multicolumn{2}{c|}{$(2.75, 3.00)$} & \multicolumn{2}{c|}{$(3.00, 3.25)$} & \multicolumn{2}{c|}{$(3.25, 3.50)$} & \multicolumn{2}{c|}{$(3.50, 4.00)$} & \multicolumn{2}{c|}{$(4.00, 4.50)$} \\ \hline

\phantom{-}$+$\phantom{-} & \phantom{$-$}1\phantom{.00} &&&&&&&&&&&&&&&& \phantom{-}\multirow{2}{*}{\begin{sideways}{$(2.00, 2.25)$}\end{sideways}}\phantom{-} \\
$-$ & 0.60 & \phantom{$-$}1\phantom{.00} &&&&&&&&&&&&&&& \\ \hline

$+$ & 0.13 & 0.48 & \phantom{$-$}1\phantom{.00} &&&&&&&&&&&&&& \multirow{2}{*}{\begin{sideways}{$(2.25, 2.50)$}\end{sideways}} \\
$-$ & 0.45 & 0.03 & 0.34 & \phantom{$-$}1\phantom{.00} &&&&&&&&&&&&& \\ \hline

$+$ & 0.26 & 0.42 & 0.48 & 0.20 & \phantom{$-$}1\phantom{.00} &&&&&&&&&&&& \multirow{2}{*}{\begin{sideways}{$(2.50, 2.75)$}\end{sideways}} \\
$-$ & 0.44 & 0.15 & 0.16 & 0.52 & 0.64 & \phantom{$-$}1\phantom{.00} & & & & & & & & & & \\ \hline

$+$ & 0.42 & 0.22 & 0.24 & 0.46 & 0.36 & 0.49 & \phantom{$-$}1\phantom{.00} &&&&&&&&&& \multirow{2}{*}{\begin{sideways}{$(2.75, 3.00)$}\end{sideways}} \\
$-$ & 0.10 & 0.50 & 0.59 & $-0.04$ & 0.48 & 0.12 & 0.51 & \phantom{$-$}1\phantom{.00} &&&&&&&&& \\ \hline

$+$ & 0.23 & 0.41 & 0.47 & 0.17 & 0.46 & 0.28 & 0.33 & 0.49 & \phantom{$-$}1\phantom{.00} &&&&&&&& \multirow{2}{*}{\begin{sideways}{$(3.00, 3.25)$}\end{sideways}} \\
$-$ & 0.45 & 0.03 & 0.02 & 0.58 & 0.21 & 0.53 & 0.47 & $-0.03$ & 0.49 & \phantom{$-$}1\phantom{.00} &&&&&&& \\ \hline

$+$ & 0.17 & 0.45 & 0.53 & 0.06 & 0.46 & 0.20 & 0.28 & 0.56 & 0.46 & 0.07 & \phantom{$-$}1\phantom{.00} &&&&&& \multirow{2}{*}{\begin{sideways}{$(3.25, 3.50)$}\end{sideways}} \\
$-$ & 0.46 & $-0.06$ & $-0.09$ & 0.61 & 0.12 & 0.53 & 0.46 & $-0.15$ & 0.09 & 0.62 & 0.19 & \phantom{$-$}1\phantom{.00} &&&&& \\ \hline

$+$ & 0.29 & 0.37 & 0.42 & 0.25 & 0.45 & 0.34 & 0.38 & 0.41 & 0.42 & 0.26 & 0.41 & 0.19 & \phantom{$-$}1\phantom{.00} &&&& \multirow{2}{*}{\begin{sideways}{$(3.50, 4.00)$}\end{sideways}} \\
$-$ & 0.43 & 0.20 & 0.21 & 0.49 & 0.35 & 0.50 & 0.48 & 0.18 & 0.31 & 0.50 & 0.24 & 0.49 & 0.66 & \phantom{$-$}1\phantom{.00} &&& \\ \hline

$+$ & 0.11 & 0.22 & 0.27 & 0.06 & 0.26 & 0.11 & 0.15 & 0.26 & 0.23 & 0.07 & 0.23 & 0.01 & 0.22 & 0.17 & \phantom{$-$}1\phantom{.00} && \multirow{2}{*}{\begin{sideways}{$(4.00, 4.50)$}\end{sideways}} \\
$-$ & 0.20 & 0.14 & 0.15 & 0.22 & 0.21 & 0.24 & 0.24 & 0.13 & 0.19 & 0.23 & 0.15 & 0.21 & 0.21 & 0.25 & 0.21 & \phantom{$-$}1\phantom{.00} & \\ \hline

& \multicolumn{1}{c}{$+$} & \multicolumn{1}{c|}{$-$} & \multicolumn{1}{c}{$+$} & \multicolumn{1}{c|}{$-$} & \multicolumn{1}{c}{$+$} & \multicolumn{1}{c|}{$-$} & \multicolumn{1}{c}{$+$} & \multicolumn{1}{c|}{$-$} & \multicolumn{1}{c}{$+$} & \multicolumn{1}{c|}{$-$} & \multicolumn{1}{c}{$+$} & \multicolumn{1}{c|}{$-$} & \multicolumn{1}{c}{$+$} & \multicolumn{1}{c|}{$-$} & \multicolumn{1}{c}{$+$} & \multicolumn{1}{c|}{$-$} & \\

\end{tabular}
\end{scriptsize}
\end{center}
\end{table}

\clearpage
\addcontentsline{toc}{section}{References}
\setboolean{inbibliography}{true}
\bibliographystyle{LHCb}
\bibliography{main,my-bibliography,LHCb-PAPER}

\ifx\mcitethebibliography\mciteundefinedmacro
\PackageError{LHCb.bst}{mciteplus.sty has not been loaded}
{This bibstyle requires the use of the mciteplus package.}\fi
\providecommand{\href}[2]{#2}
\begin{mcitethebibliography}{10}
\mciteSetBstSublistMode{n}
\mciteSetBstMaxWidthForm{subitem}{\alph{mcitesubitemcount})}
\mciteSetBstSublistLabelBeginEnd{\mcitemaxwidthsubitemform\space}
{\relax}{\relax}

\bibitem{DY-NNLO1}
P.~Rijken and W.~van Neerven, \ifthenelse{\boolean{articletitles}}{{\it {Order
  $\alpha_{s}^{2}$ contributions to the Drell-Yan cross-section at fixed target
  energies}}, }{}\href{http://dx.doi.org/10.1103/PhysRevD.51.44}{Phys.\ Rev.\
  {\bf D51} (1995) 44}, \href{http://arxiv.org/abs/hep-ph/9408366}{{\tt
  arXiv:hep-ph/9408366}}\relax
\mciteBstWouldAddEndPuncttrue
\mciteSetBstMidEndSepPunct{\mcitedefaultmidpunct}
{\mcitedefaultendpunct}{\mcitedefaultseppunct}\relax
\EndOfBibitem
\bibitem{DY-NNLO2}
R.~Hamberg, W.~van Neerven, and T.~Matsuura,
  \ifthenelse{\boolean{articletitles}}{{\it {A complete calculation of the
  order $\alpha_{s}^{2}$ correction to the Drell-Yan $K$-factor}},
  }{}\href{http://dx.doi.org/10.1016/0550-3213(91)90064-5}{Nucl.\ Phys.\  {\bf
  B359} (1991) 343}\relax
\mciteBstWouldAddEndPuncttrue
\mciteSetBstMidEndSepPunct{\mcitedefaultmidpunct}
{\mcitedefaultendpunct}{\mcitedefaultseppunct}\relax
\EndOfBibitem
\bibitem{DY-NNLO3}
W.~van Neerven and E.~Zijlstra, \ifthenelse{\boolean{articletitles}}{{\it {The
  O$(\alpha_{s}^{2})$ corrected Drell-Yan $K$-factor in the DIS and MS
  scheme}}, }{}\href{http://dx.doi.org/10.1016/0550-3213(92)90078-P}{Nucl.\
  Phys.\  {\bf B382} (1992) 11}\relax
\mciteBstWouldAddEndPuncttrue
\mciteSetBstMidEndSepPunct{\mcitedefaultmidpunct}
{\mcitedefaultendpunct}{\mcitedefaultseppunct}\relax
\EndOfBibitem
\bibitem{DY-NNLO4}
R.~V. Harlander and W.~B. Kilgore, \ifthenelse{\boolean{articletitles}}{{\it
  {Next-to-next-to-leading order Higgs production at hadron colliders}},
  }{}\href{http://dx.doi.org/10.1103/PhysRevLett.88.201801}{Phys.\ Rev.\ Lett.\
   {\bf 88} (2002) 201801}, \href{http://arxiv.org/abs/hep-ph/0201206}{{\tt
  arXiv:hep-ph/0201206}}\relax
\mciteBstWouldAddEndPuncttrue
\mciteSetBstMidEndSepPunct{\mcitedefaultmidpunct}
{\mcitedefaultendpunct}{\mcitedefaultseppunct}\relax
\EndOfBibitem
\bibitem{DY-NNLO5}
C.~Anastasiou, L.~J. Dixon, K.~Melnikov, and F.~Petriello,
  \ifthenelse{\boolean{articletitles}}{{\it {High-precision QCD at hadron
  colliders: Electroweak gauge boson rapidity distributions at next-to-next-to
  leading order}},
  }{}\href{http://dx.doi.org/10.1103/PhysRevD.69.094008}{Phys.\ Rev.\  {\bf
  D69} (2004) 094008}, \href{http://arxiv.org/abs/hep-ph/0312266}{{\tt
  arXiv:hep-ph/0312266}}\relax
\mciteBstWouldAddEndPuncttrue
\mciteSetBstMidEndSepPunct{\mcitedefaultmidpunct}
{\mcitedefaultendpunct}{\mcitedefaultseppunct}\relax
\EndOfBibitem
\bibitem{PDF-MSTW}
R.~Thorne, A.~Martin, W.~Stirling, and G.~Watt,
  \ifthenelse{\boolean{articletitles}}{{\it {Parton distributions and QCD at
  \lhcb}}, }{}\href{http://arxiv.org/abs/0808.1847}{{\tt
  arXiv:0808.1847}}\relax
\mciteBstWouldAddEndPuncttrue
\mciteSetBstMidEndSepPunct{\mcitedefaultmidpunct}
{\mcitedefaultendpunct}{\mcitedefaultseppunct}\relax
\EndOfBibitem
\bibitem{LHCb-PAPER-2012-008}
LHCb collaboration, R.~Aaij {\em et~al.},
  \ifthenelse{\boolean{articletitles}}{{\it {Inclusive $W$ and $Z$ production
  in the forward region at $\sqrt{s}=7$ TeV}},
  }{}\href{http://dx.doi.org/10.1007/JHEP06(2012)058}{JHEP {\bf 06} (2012)
  058}, \href{http://arxiv.org/abs/1204.1620}{{\tt arXiv:1204.1620}}\relax
\mciteBstWouldAddEndPuncttrue
\mciteSetBstMidEndSepPunct{\mcitedefaultmidpunct}
{\mcitedefaultendpunct}{\mcitedefaultseppunct}\relax
\EndOfBibitem
\bibitem{PDF-ABM12}
S.~Alekhin, J.~Bluemlein, and S.~Moch,
  \ifthenelse{\boolean{articletitles}}{{\it {The ABM parton distributions tuned
  to LHC data}}, }{}\href{http://dx.doi.org/10.1103/PhysRevD.89.054028}{Phys.\
  Rev.\  {\bf D89} (2014) 054028}, \href{http://arxiv.org/abs/1310.3059}{{\tt
  arXiv:1310.3059}}\relax
\mciteBstWouldAddEndPuncttrue
\mciteSetBstMidEndSepPunct{\mcitedefaultmidpunct}
{\mcitedefaultendpunct}{\mcitedefaultseppunct}\relax
\EndOfBibitem
\bibitem{PDF-NNPDF23}
R.~D. Ball {\em et~al.}, \ifthenelse{\boolean{articletitles}}{{\it {Parton
  distributions with \lhc data}},
  }{}\href{http://dx.doi.org/10.1016/j.nuclphysb.2012.10.003}{Nucl.\ Phys.\
  {\bf B867} (2013) 244}, \href{http://arxiv.org/abs/1207.1303}{{\tt
  arXiv:1207.1303}}\relax
\mciteBstWouldAddEndPuncttrue
\mciteSetBstMidEndSepPunct{\mcitedefaultmidpunct}
{\mcitedefaultendpunct}{\mcitedefaultseppunct}\relax
\EndOfBibitem
\bibitem{ATLAS}
ATLAS collaboration, G.~Aad {\em et~al.},
  \ifthenelse{\boolean{articletitles}}{{\it {Measurement of the inclusive
  $W^\pm$ and $Z/\gamma^{*}$ cross sections in the electron and muon decay
  channels in $pp$ collisions at $\sqrt{s}=7$ TeV with the ATLAS detector}},
  }{}\href{http://dx.doi.org/10.1103/PhysRevD.85.072004}{Phys.\ Rev.\  {\bf
  D85} (2012) 072004}, \href{http://arxiv.org/abs/1109.5141}{{\tt
  arXiv:1109.5141}}\relax
\mciteBstWouldAddEndPuncttrue
\mciteSetBstMidEndSepPunct{\mcitedefaultmidpunct}
{\mcitedefaultendpunct}{\mcitedefaultseppunct}\relax
\EndOfBibitem
\bibitem{CMS}
CMS collaboration, S.~Chatrchyan {\em et~al.},
  \ifthenelse{\boolean{articletitles}}{{\it {Measurement of the muon charge
  asymmetry in inclusive $pp \to W + X$ production at $\sqrt{s}$=7 TeV and an
  improved determination of light parton distribution functions}},
  }{}\href{http://arxiv.org/abs/1312.6283}{{\tt arXiv:1312.6283}}\relax
\mciteBstWouldAddEndPuncttrue
\mciteSetBstMidEndSepPunct{\mcitedefaultmidpunct}
{\mcitedefaultendpunct}{\mcitedefaultseppunct}\relax
\EndOfBibitem
\bibitem{Alves:2008zz}
LHCb collaboration, A.~A. Alves~Jr. {\em et~al.},
  \ifthenelse{\boolean{articletitles}}{{\it {The \lhcb detector at the LHC}},
  }{}\href{http://dx.doi.org/10.1088/1748-0221/3/08/S08005}{JINST {\bf 3}
  (2008) S08005}\relax
\mciteBstWouldAddEndPuncttrue
\mciteSetBstMidEndSepPunct{\mcitedefaultmidpunct}
{\mcitedefaultendpunct}{\mcitedefaultseppunct}\relax
\EndOfBibitem
\bibitem{LUMI-MEER}
S.~van~der Meer, \ifthenelse{\boolean{articletitles}}{{\it {Calibration of the
  effective beam height in the ISR}}, }{}CERN-ISR-PO {\bf 68-31} (1968)\relax
\mciteBstWouldAddEndPuncttrue
\mciteSetBstMidEndSepPunct{\mcitedefaultmidpunct}
{\mcitedefaultendpunct}{\mcitedefaultseppunct}\relax
\EndOfBibitem
\bibitem{FerroLuzzi:2005em}
M.~Ferro-Luzzi, \ifthenelse{\boolean{articletitles}}{{\it {Proposal for an
  absolute luminosity determination in colliding beam experiments using vertex
  detection of beam-gas interactions}},
  }{}\href{http://dx.doi.org/10.1016/j.nima.2005.07.010}{Nucl.\ Instrum.\
  Meth.\  {\bf A553} (2005) 388}\relax
\mciteBstWouldAddEndPuncttrue
\mciteSetBstMidEndSepPunct{\mcitedefaultmidpunct}
{\mcitedefaultendpunct}{\mcitedefaultseppunct}\relax
\EndOfBibitem
\bibitem{LHCB-PAPER-2014-047}
LHCb collaboration, R.~Aaij {\em et~al.},
  \ifthenelse{\boolean{articletitles}}{{\it {Precision luminosity measurements
  at LHCb}}, }{}\href{http://dx.doi.org/10.1088/1748-0221/9/12/P12005}{JINST
  {\bf 9} (2014) P12005}, \href{http://arxiv.org/abs/1410.0149}{{\tt
  arXiv:1410.0149}}\relax
\mciteBstWouldAddEndPuncttrue
\mciteSetBstMidEndSepPunct{\mcitedefaultmidpunct}
{\mcitedefaultendpunct}{\mcitedefaultseppunct}\relax
\EndOfBibitem
\bibitem{Sjostrand:2006za}
T.~Sj\"{o}strand, S.~Mrenna, and P.~Skands,
  \ifthenelse{\boolean{articletitles}}{{\it {PYTHIA 6.4 physics and manual}},
  }{}\href{http://dx.doi.org/10.1088/1126-6708/2006/05/026}{JHEP {\bf 05}
  (2006) 026}, \href{http://arxiv.org/abs/hep-ph/0603175}{{\tt
  arXiv:hep-ph/0603175}}\relax
\mciteBstWouldAddEndPuncttrue
\mciteSetBstMidEndSepPunct{\mcitedefaultmidpunct}
{\mcitedefaultendpunct}{\mcitedefaultseppunct}\relax
\EndOfBibitem
\bibitem{LHCb-PROC-2010-056}
I.~Belyaev {\em et~al.}, \ifthenelse{\boolean{articletitles}}{{\it {Handling of
  the generation of primary events in GAUSS, the \lhcb simulation framework}},
  }{}\href{http://dx.doi.org/10.1109/NSSMIC.2010.5873949}{Nuclear Science
  Symposium Conference Record (NSS/MIC) {\bf IEEE} (2010) 1155}\relax
\mciteBstWouldAddEndPuncttrue
\mciteSetBstMidEndSepPunct{\mcitedefaultmidpunct}
{\mcitedefaultendpunct}{\mcitedefaultseppunct}\relax
\EndOfBibitem
\bibitem{Lange:2001uf}
D.~J. Lange, \ifthenelse{\boolean{articletitles}}{{\it {The EvtGen particle
  decay simulation package}},
  }{}\href{http://dx.doi.org/10.1016/S0168-9002(01)00089-4}{Nucl.\ Instrum.\
  Meth.\  {\bf A462} (2001) 152}\relax
\mciteBstWouldAddEndPuncttrue
\mciteSetBstMidEndSepPunct{\mcitedefaultmidpunct}
{\mcitedefaultendpunct}{\mcitedefaultseppunct}\relax
\EndOfBibitem
\bibitem{Golonka:2005pn}
P.~Golonka and Z.~Was, \ifthenelse{\boolean{articletitles}}{{\it {PHOTOS Monte
  Carlo: a precision tool for QED corrections in $Z$ and $W$ decays}},
  }{}\href{http://dx.doi.org/10.1140/epjc/s2005-02396-4}{Eur.\ Phys.\ J.\  {\bf
  C45} (2006) 97}, \href{http://arxiv.org/abs/hep-ph/0506026}{{\tt
  arXiv:hep-ph/0506026}}\relax
\mciteBstWouldAddEndPuncttrue
\mciteSetBstMidEndSepPunct{\mcitedefaultmidpunct}
{\mcitedefaultendpunct}{\mcitedefaultseppunct}\relax
\EndOfBibitem
\bibitem{Allison:2006ve}
Geant4 collaboration, J.~Allison {\em et~al.},
  \ifthenelse{\boolean{articletitles}}{{\it {Geant4 developments and
  applications}}, }{}\href{http://dx.doi.org/10.1109/TNS.2006.869826}{IEEE
  Trans.\ Nucl.\ Sci.\  {\bf 53} (2006) 270}\relax
\mciteBstWouldAddEndPuncttrue
\mciteSetBstMidEndSepPunct{\mcitedefaultmidpunct}
{\mcitedefaultendpunct}{\mcitedefaultseppunct}\relax
\EndOfBibitem
\bibitem{Agostinelli:2002hh}
Geant4 collaboration, S.~Agostinelli {\em et~al.},
  \ifthenelse{\boolean{articletitles}}{{\it {Geant4: a simulation toolkit}},
  }{}\href{http://dx.doi.org/10.1016/S0168-9002(03)01368-8}{Nucl.\ Instrum.\
  Meth.\  {\bf A506} (2003) 250}\relax
\mciteBstWouldAddEndPuncttrue
\mciteSetBstMidEndSepPunct{\mcitedefaultmidpunct}
{\mcitedefaultendpunct}{\mcitedefaultseppunct}\relax
\EndOfBibitem
\bibitem{LHCb-PROC-2011-006}
M.~Clemencic {\em et~al.}, \ifthenelse{\boolean{articletitles}}{{\it {The \lhcb
  simulation application, GAUSS: design, evolution and experience}},
  }{}\href{http://dx.doi.org/10.1088/1742-6596/331/3/032023}{{J.\ Phys.\ Conf.\
  Ser.\ } {\bf 331} (2011) 032023}\relax
\mciteBstWouldAddEndPuncttrue
\mciteSetBstMidEndSepPunct{\mcitedefaultmidpunct}
{\mcitedefaultendpunct}{\mcitedefaultseppunct}\relax
\EndOfBibitem
\bibitem{GEN-RESBOS1}
G.~A. Ladinsky and C.-P. Yuan, \ifthenelse{\boolean{articletitles}}{{\it {The
  nonperturbative regime in QCD resummation for gauge boson production at
  hadron colliders}},
  }{}\href{http://dx.doi.org/10.1103/PhysRevD.50.R4239}{Phys.\ Rev.\  {\bf D50}
  (1994) 4239}, \href{http://arxiv.org/abs/hep-ph/9311341}{{\tt
  arXiv:hep-ph/9311341}}\relax
\mciteBstWouldAddEndPuncttrue
\mciteSetBstMidEndSepPunct{\mcitedefaultmidpunct}
{\mcitedefaultendpunct}{\mcitedefaultseppunct}\relax
\EndOfBibitem
\bibitem{GEN-RESBOS2}
C.~Balazs and C.-P. Yuan, \ifthenelse{\boolean{articletitles}}{{\it {Soft gluon
  effects on lepton pairs at hadron colliders}},
  }{}\href{http://dx.doi.org/10.1103/PhysRevD.56.5558}{Phys.\ Rev.\  {\bf D56}
  (1997) 5558}, \href{http://arxiv.org/abs/hep-ph/9704258}{{\tt
  arXiv:hep-ph/9704258}}\relax
\mciteBstWouldAddEndPuncttrue
\mciteSetBstMidEndSepPunct{\mcitedefaultmidpunct}
{\mcitedefaultendpunct}{\mcitedefaultseppunct}\relax
\EndOfBibitem
\bibitem{GEN-RESBOS3}
F.~Landry, R.~Brock, P.~M. Nadolsky, and C.-P. Yuan,
  \ifthenelse{\boolean{articletitles}}{{\it {Fermilab \tevatron run-1 \Z boson
  data and Collins-Soper-Sterman resummation formalism}},
  }{}\href{http://dx.doi.org/10.1103/PhysRevD.67.073016}{Phys.\ Rev.\  {\bf
  D67} (2003) 073016}, \href{http://arxiv.org/abs/hep-ph/0212159}{{\tt
  arXiv:hep-ph/0212159}}\relax
\mciteBstWouldAddEndPuncttrue
\mciteSetBstMidEndSepPunct{\mcitedefaultmidpunct}
{\mcitedefaultendpunct}{\mcitedefaultseppunct}\relax
\EndOfBibitem
\bibitem{PDF-CT10}
J.~Gao {\em et~al.}, \ifthenelse{\boolean{articletitles}}{{\it {CT10
  next-to-next-to-leadingorder global analysis of QCD}},
  }{}\href{http://dx.doi.org/10.1103/PhysRevD.89.033009}{Phys.\ Rev.\  {\bf
  D89} (2014) 033009}, \href{http://arxiv.org/abs/1302.6246}{{\tt
  arXiv:1302.6246}}\relax
\mciteBstWouldAddEndPuncttrue
\mciteSetBstMidEndSepPunct{\mcitedefaultmidpunct}
{\mcitedefaultendpunct}{\mcitedefaultseppunct}\relax
\EndOfBibitem
\bibitem{GEN-FEWZ2}
R.~Gavin, Y.~Li, F.~Petriello, and S.~Quackenbush,
  \ifthenelse{\boolean{articletitles}}{{\it {FEWZ 2.0: a code for hadronic \Z
  production at next-to-next-to-leading order}},
  }{}\href{http://dx.doi.org/10.1016/j.cpc.2011.06.008}{Comput.\ Phys.\
  Commun.\  {\bf 182} (2011) 2388}, \href{http://arxiv.org/abs/1011.3540}{{\tt
  arXiv:1011.3540}}\relax
\mciteBstWouldAddEndPuncttrue
\mciteSetBstMidEndSepPunct{\mcitedefaultmidpunct}
{\mcitedefaultendpunct}{\mcitedefaultseppunct}\relax
\EndOfBibitem
\bibitem{GEN-FEWZ3}
Y.~Li and F.~Petriello, \ifthenelse{\boolean{articletitles}}{{\it {Combining
  QCD and electroweak corrections to dilepton production in the framework of
  the FEWZ simulation code}},
  }{}\href{http://dx.doi.org/10.1103/PhysRevD.86.094034}{Phys.\ Rev.\  {\bf
  D86} (2012) 094034}, \href{http://arxiv.org/abs/1208.5967}{{\tt
  arXiv:1208.5967}}\relax
\mciteBstWouldAddEndPuncttrue
\mciteSetBstMidEndSepPunct{\mcitedefaultmidpunct}
{\mcitedefaultendpunct}{\mcitedefaultseppunct}\relax
\EndOfBibitem
\bibitem{PDF-HERA15}
ZEUS collaboration, H1 collaboration, A.~Cooper-Sarkar,
  \ifthenelse{\boolean{articletitles}}{{\it {PDF fits at HERA}}, }{}PoS {\bf
  EPS-HEP2011} (2011) 320, \href{http://arxiv.org/abs/1112.2107}{{\tt
  arXiv:1112.2107}}\relax
\mciteBstWouldAddEndPuncttrue
\mciteSetBstMidEndSepPunct{\mcitedefaultmidpunct}
{\mcitedefaultendpunct}{\mcitedefaultseppunct}\relax
\EndOfBibitem
\bibitem{PDF-JR09}
P.~Jimenez-Delgado and E.~Reya, \ifthenelse{\boolean{articletitles}}{{\it
  {Dynamical next-to-next-to-leading order parton distributions}},
  }{}\href{http://dx.doi.org/10.1103/PhysRevD.79.074023}{Phys.\ Rev.\  {\bf
  D79} (2009) 074023}, \href{http://arxiv.org/abs/0810.4274}{{\tt
  arXiv:0810.4274}}\relax
\mciteBstWouldAddEndPuncttrue
\mciteSetBstMidEndSepPunct{\mcitedefaultmidpunct}
{\mcitedefaultendpunct}{\mcitedefaultseppunct}\relax
\EndOfBibitem
\bibitem{PDF-MSTW08}
A.~D. Martin, W.~J. Stirling, R.~S. Thorne, and G.~Watt,
  \ifthenelse{\boolean{articletitles}}{{\it {Parton distributions for the
  \lhc}}, }{}\href{http://dx.doi.org/10.1140/epjc/s10052-009-1072-5}{Eur.\
  Phys.\ J.\  {\bf C63} (2009) 189}, \href{http://arxiv.org/abs/0901.0002}{{\tt
  arXiv:0901.0002}}\relax
\mciteBstWouldAddEndPuncttrue
\mciteSetBstMidEndSepPunct{\mcitedefaultmidpunct}
{\mcitedefaultendpunct}{\mcitedefaultseppunct}\relax
\EndOfBibitem
\bibitem{TFRACTIONFITTER}
R.~J. Barlow and C.~Beeston, \ifthenelse{\boolean{articletitles}}{{\it {Fitting
  using finite Monte Carlo samples}},
  }{}\href{http://dx.doi.org/10.1016/0010-4655(93)90005-W}{Comput.\ Phys.\
  Commun.\  {\bf 77} (1993) 219}\relax
\mciteBstWouldAddEndPuncttrue
\mciteSetBstMidEndSepPunct{\mcitedefaultmidpunct}
{\mcitedefaultendpunct}{\mcitedefaultseppunct}\relax
\EndOfBibitem
\bibitem{LHCb-PAPER-2012-036}
LHCb collaboration, R.~Aaij {\em et~al.},
  \ifthenelse{\boolean{articletitles}}{{\it {Measurement of the cross-section
  for $Z \to e^+e^-$ production in $pp$ collisions at $\sqrt{s}=7$ TeV}},
  }{}\href{http://dx.doi.org/10.1007/JHEP02(2013)106}{JHEP {\bf 02} (2013)
  106}, \href{http://arxiv.org/abs/1212.4620}{{\tt arXiv:1212.4620}}\relax
\mciteBstWouldAddEndPuncttrue
\mciteSetBstMidEndSepPunct{\mcitedefaultmidpunct}
{\mcitedefaultendpunct}{\mcitedefaultseppunct}\relax
\EndOfBibitem
\end{mcitethebibliography}

\clearpage
\centerline{\large\bf LHCb collaboration}
\begin{flushleft}
\small
R.~Aaij$^{41}$, 
B.~Adeva$^{37}$, 
M.~Adinolfi$^{46}$, 
A.~Affolder$^{52}$, 
Z.~Ajaltouni$^{5}$, 
S.~Akar$^{6}$, 
J.~Albrecht$^{9}$, 
F.~Alessio$^{38}$, 
M.~Alexander$^{51}$, 
S.~Ali$^{41}$
% ...

{\footnotesize \it
% ...
$ ^{5}$Clermont Universit\'{e}, Universit\'{e} Blaise Pascal, CNRS/IN2P3, LPC, Clermont-Ferrand, France\\
$ ^{6}$CPPM, Aix-Marseille Universit\'{e}, CNRS/IN2P3, Marseille, France\\
% ...
$ ^{9}$Fakult\"{a}t Physik, Technische Universit\"{a}t Dortmund, Dortmund, Germany\\
% ...
$ ^{37}$Universidad de Santiago de Compostela, Santiago de Compostela, Spain\\
$ ^{38}$European Organization for Nuclear Research (CERN), Geneva, Switzerland\\
% ...
$ ^{41}$Nikhef National Institute for Subatomic Physics, Amsterdam, The Netherlands\\
% ...
$ ^{46}$H.H. Wills Physics Laboratory, University of Bristol, Bristol, United Kingdom\\
% ...
$ ^{51}$School of Physics and Astronomy, University of Glasgow, Glasgow, United Kingdom\\
$ ^{52}$Oliver Lodge Laboratory, University of Liverpool, Liverpool, United Kingdom\\
% ...
}
\end{flushleft}

\end{document}